\documentstyle[graphics,onecolumn]{mn}
\newcommand{\mincir}{\raise
-2.truept\hbox{\rlap{\hbox{$\sim$}}\raise5.truept 
\hbox{$<$}\ }}
\newcommand{\magcir}{\raise
-2.truept\hbox{\rlap{\hbox{$\sim$}}\raise5.truept
\hbox{$>$}\ }}
\newcommand{\minmag}{\raise-2.truept\hbox{\rlap{\hbox{$<$}}\raise
6.truept\hbox
{$>$}\ }}

\newcommand{\be}{\begin{equation}}
\newcommand{\ee}{\end{equation}}
\newcommand{\ba}{\begin{eqnarray}}
\newcommand{\ea}{\end{eqnarray}}
\newcommand{\brr}{\begin{array}}
 
\newcommand{\err}{\end{array}}
\newcommand{\bc}{\begin{center}}
\newcommand{\ec}{\end{center}}

\newcommand{\hm}{\,h^{-1}{\rm Mpc}}

\title[Probing the Intergalactic Medium with the Ly$\alpha$ forest
along multiple lines of sight] {Probing the Intergalactic Medium with
the Ly$\alpha$ forest along multiple lines of sight to distant QSOs}
\author[M. Viel, S. Matarrese, H. J. Mo, M. G. Haehnelt \& T. Theuns] {M. Viel $^{1, 2}$,
S. Matarrese $^{1, 2}$, H. J. Mo $^2$, M. G. Haehnelt $^3$ \& Tom
Theuns $^{4}$ \\ $^1$ Dipartimento di Fisica `Galileo Galilei', via
Marzolo 8, I-35131 Padova, Italy \\ $^2$ Max-Planck-Institut f\"ur
Astrophysik, Karl-Schwarzschild-Strasse 1, D-85741 Garching, Germany\\
$^3$ Astrophysics Group, Imperial College of Science, Technology and
Medicine, Prince Consort Road, London SW7 2BW, UK\\ $^4$ Institute of
Astronomy, Madingley Road, Cambridge CB3 0HA, UK.\\
Email: viel@pd.infn.it, matarrese@pd.infn.it, hom@mpa-garching.mpg.de, 
m.haehnelt@ic.ac.uk, tt@ast.cam.ac.uk}

\date{Accepted for publication in MNRAS 2001 October 18}

\begin{document}

\maketitle

\begin{abstract}
We present an effective implementation of analytical calculations of
the Ly$\alpha$ opacity distribution of the Intergalactic Medium (IGM)
along multiple lines of sight (LOS) to distant quasars in a
cosmological setting.  The method assumes that the distribution of
neutral hydrogen follows that of an underlying dark matter density
field and that the density distribution is a (local) lognormal
distribution. It fully accounts for the expected correlations between
LOS and the cosmic variance in the large-scale modes of the dark
matter distribution. Strong correlations extending up to $\sim 300$
kpc (proper) and more are found at redshift $z\sim 2 \to 3$, in
agreement with observations. These correlations are investigated using
the cross-correlation coefficient and the cross-power spectrum of the
flux distribution along different LOS and by identifying coincident
absorption features as fitted with a Voigt profile fitting
routine. The cross-correlation coefficient between the LOS can be used
to constrain the shape-parameter $\Gamma$ of the power spectrum if the
temperature and the temperature density relation of the IGM can be
determined indepedently.  We also propose a new technique to recover
the 3D linear dark matter power spectrum by integrating over 1D flux
cross-spectra which is complementary to the usual `differentiation' of
1D auto-spectra. The cross-power spectrum suffers much less from
errors uncorrelated in different LOS, like those introduced by
continuum fitting. Investigations of the flux correlations in adjacent
LOS should thus allow to extend studies of the dark matter power
spectrum with the Ly$\alpha$ forest to significantly larger scales
than is possible with flux auto-power spectra. 30 pairs with
separation of \linebreak 1-2 arcmin should be sufficient to determine the 1D
cross-spectrum at scales of \linebreak$60\, h^{-1}$ Mpc with an
accuracy of about 30\% (corresponding to a 15\% error of the {\it rms}
density fluctuation amplitude) if the error is dominated by cosmic
variance.
\end{abstract}


\section{Introduction}
 The current understanding of QSO spectra blueward of
Ly$\alpha$ emission, the so-called Ly$\alpha$ forest, is based on
the idea that the Ly$\alpha$ absorption is produced by the
inhomogenous distribution of the Intergalactic Medium (IGM) along the
line of sight (Bahcall \& Salpeter 1965; Gunn \& Peterson
1965). The IGM is thereby  believed to be warm ($\sim 10^{4}$ K) and
photoionized. The rather high flux of the ultraviolet background
radiation results in a small neutral hydrogen fraction with
Ly$\alpha$ optical depth of order unity  which is responsible for a
`fluctuating Gunn-Peterson effect'  (see e.g. Rauch 1998, for a
review). Such a fluctuating Gunn-Peterson effect arises
naturally  in standard hierarchical models for structure formation
where the matter clusters gravitationally into filamentary and
sheet-like structures.  The low-column density (${\rm N}_{\rm HI}
\leq 10^{14.5}$ cm$^{-2}$)  absorption lines are generated by local
fluctuations in the IGM, which smoothly trace the mildly non-linear
dark matter filaments  and sheets on scales larger the Jeans scale
of the  photoionized IGM (Cen {\it et al.} 1994, Miralda-Escud\'e {\it
et al.} 1996, Zhang {\it et al.} 1998). 

This picture is supported by analytical studies based on simple models
for the IGM dynamics.  Various models of this kind have been proposed,
based on either a local non-linear mapping of the linear density
contrast, such as the lognormal model (Coles \& Jones 1991), applied
to the IGM dynamics (Bi, B\"orner \& Chu 1992; Bi 1993; Bi, Ge \& Fang
1995; Bi \& Davidsen 1997, hereafter BD97), or on suitable
modifications of the Zel'dovich approximation (Zel'dovich 1970) to
account for the smoothing caused by gas pressure on the baryon Jeans
scale (Reisenegger \& Miralda-Escud\'e 1995; Gnedin \& Hui 1996; Hui,
Gnedin \& Zhang 1997, Matarrese \& Mohayaee 2001). 

The most convincing support for this picture comes, however, from the
comparison of simulated spectra produced from hydrodynamical numerical
simulations with observed spectra (Cen {\it et al.} 1994; Zhang,
Anninos \& Norman 1995, 1997; Miralda-Escud\'e {\it et al.} 1996;
Hernquist {\it et al.}  1996; Charlton {\it et al.}  1997; Theuns {\it
et al.} 1998).  The numerical simulations have been demonstrated to
reproduce many observed properties of the Ly$\alpha$ forest very well.
Simple analytic schemes, as the one developed here, can be calibrated
by the results of numerical simulations. They then become an important
complementary tool for studying the Ly$\alpha$ forest. They can be
used to explore larger regions of model parameter space and can better
account for the cosmic variance of large-scale modes.  These are
poorly probed by existing hydro-simulations which have to adopt
relatively small computational boxes in order to resolve the Jeans
scale of the warm photoionized IGM.  

Observationally, the unprecedented high resolution observations of the
Keck HIRES spectrograph and the UV spectroscopic capabilities of the
HST have been instrumental in shaping our current understanding of
Ly$\alpha$ forest.  HIRES allowed to detect lines with column
densities as low as ${\rm N}_{\rm HI} \sim 10^{12}$ cm$^{-2}$ while
HST made a detailed analysis of the low-redshift Ly$\alpha$ forest at
$z<1.6$ possible.  From the study of absorption spectra along single
lines of sight (LOS) to distant QSOs we have, for example, gained important
information on the baryon density of the Universe (Rauch {\it et al.} 1997)
and on the temperature and equation of state of the IGM (Schaye {\it
et al.}  2000). Another important application is the determination of
shape and amplitude of the power spectrum of the spatial distribution
of dark matter at redshift $z\sim 3$, from the fluctuating Ly$\alpha$
flux, which places important constraints on the parameters of
structure formation models (Croft {\it et al.} 1998, 1999, 2000;
Nusser \& Haehnelt 1999, 2000; White \& Croft 2000; Narayanan {\it et
al.}  2000). 

In this paper, we will concentrate on the information which can be
extracted from the expected flux-correlations in adjacent LOS (see
Charlton {\it et al.}  1997 for an analysis of hydro simulations).
Observations of multiple systems are an excellent tool to probe the
actual 3D distribution of matter in the Universe and to give estimates
of the size of the absorbing structurs. Another important application
of this type of study is to constrain the global geometry of the
Universe (Hui, Stebbins \& Burles (1999) and McDonald \&
Miralda-Escud\'e (1999)).

There is a number of cases in which common absorption systems in
spatially separated LOS have been observed. These are either multiple
images of gravitationally lensed quasars (Foltz {\it et al.} 1984;
Smette {\it et al.} 1992, 1995; Rauch {\it et al.} 1999) or close
quasar pairs (Bechtold {\it et al.} 1994; Dinshaw {\it et al.}  1995,
1995, 1997; Fang {\it et al.} 1996; Crotts \& Fang 1998; D'Odorico {\it et
al.} 1998; Petitjean {\it et al.} 1998; Williger {\it et al.}  2000;
Liske {\it et al.}  2000).  The results concerning the typical size of
the absorbing structures are somewhat controversial. Crotts \& Fang
(1998) analysed a total number of five QSOs in close groupings: a pair
and a triplet, using Keck and HST data, with different separations
ranging from 9.5 to 177 arcsec (corresponding to a proper distance of
40-700 $h^{-1}$ kpc in an Einstein-de Sitter Universe) in a redshift range
$0.48<z<2.52$.  For the strongest lines identified by Voigt profile
fitting they found a tight correspondence between lines in different
LOS up to a proper separation of $0.5-0.8 \hm$.  Their estimate of the
size of the absorbers using a Bayesian model (Fang {\it et al.} 1996)
with the assumption that the absorbers are spherical with uniform
radius did, however, show a dependence on the separation of the QSO
pair.  This suggests that the assumption of a spherical absorber is
not correct, and that the absorbers are elongated or sheet-like
(Charlton {\it et al.} 1997; see also Rauch \& Haehnelt 1995, for an
independent argument).  A similar analysis by D'Odorico {\it et al.}
(1998) of several QSO pairs with a median redshift of $z=2.13$, gave a
radius of a few hundred kpc.  D'Odorico {\it et al.} used the same
Bayesian model as Crott \& Fang and found it impossible to distinguish
between a population of disk-like absorbers and a population of
spherical clouds with different radii. Petitjean {\it et al.}  (1998)
analysed HST observations of a QSO pair over a redshift range
$0.833<z<1.438$, and obtained a typical size of the Ly$\alpha$
absorber of $500h^{-1}$ kpc. Liske {\it et al.}  (2000) investigated a
system of 10 QSOs concentrated in a field of 1-deg$^2$ over the
redshift range $2.2<z<3.4$. They found correlations across lines of
sight with proper separation $< 3 h^{-1}$ Mpc. Williger {\it et al.}
(2000) investigated a grouping of 10 QSOs in the redshift range
$2.15<z<3.37$ and found a correlation length up to 26 $h^{-1}$
comoving Mpc.  More recently, Young {\it et al.} (2000) have analysed
a triple system and have found a coherence length of $0.5-1$ Mpc for a
redshift range $0.4<z<0.9$.

Here we will use an analytical method to calculate the Ly$\alpha$
opacity distribution of the IGM along multiple
LOS to distant quasars in a cosmological setting.
From these we calculate absorption spectra with varying transverse
separation between LOS pairs. We then use the cross-correlation
coefficient, as a measure of the characteristic size of the absorber,
which better describes the complicated geometrical structure of the
absorbers suggested by numerical simulations.  We further investigate
the virtues of the flux cross-power spectrum in constraining the
underlying mass density field.  To make connections with the
observational studies mentioned above we also perform an analysis of
coincident absorption lines as identified with the Voigt profile
routine AUTOVP, kindly provided by Romeel Dav\'e. 

The plan of the paper is as follows. Section 2 presents the lognormal
model for IGM dynamics and describes the algorithm which allows us to
simulate spatially correlated LOS through the Ly$\alpha$ forest. In
Section 3 we give the relations which are used to simulate the
Ly$\alpha$ flux from the IGM local density and peculiar velocity
fields and we use the cross-correlation coefficient and the
cross-power spectra to quantify the flux correlations. In Section 4 we
perform the coincidence analysis of absorption lines fitted with the
Voigt-profile fitting procedure. In Section 5 we propose a new
procedure for recovering the 3D dark matter power spectrum by
integrating the 1D cross-spectra over the transverse separation and we
show the main advantages in using the cross-spectra information.
Section 6 contains a discussion and our conclusions.
\section{Method}

\subsection{The lognormal model of the IGM}

We implement here the model introduced by Bi and collaborators (Bi
{\it et al.} 1992, 1995; Bi 1993; BD97), to simulate low
column-density Ly$\alpha$ absorption systems along the LOS, which we
then extend to simulate multiple LOS to distant QSOs. This simple
model predicts many properties, such as the column density
distribution function and the distribution of the b-parameter, which
can be directly compared with observations (BD97). Recently, the BD97
model has been used by Roy Choudhury {\it et al.} (2000, 2001) to
study neutral hydrogen correlation functions along and transverse to
LOS. Feng \& Fang (2000) also adopted the BD97 method to analyse
non-Gaussian effects in the Ly$\alpha$ transmitted flux stressing
their importance for the reconstruction of the initial mass density
field.

The BD97 model is based on the assumption that the low-column density 
Ly$\alpha$ forest is produced by smooth fluctuations 
in the intergalactic medium which  arise as a result of 
gravitational instability. 
Linear density perturbations of the intergalactic medium 
$\delta_0^{\rm IGM}({\bf x}, z)$, can be related to DM linear
overdensities by a convolution.  
In Fourier space one usually assumes 
\be
\delta_0^{\rm IGM} ({\bf k}, z) = {\delta_0^{\rm DM}
({\bf k}, z) \over 1 + k^2/k_J^2(z)} \equiv 
W_{\rm IGM}(k,z) D_+(z) \delta_0^{\rm DM}({\bf k})
\ee
where $D_+(z)$ is the linear growing mode of dark matter density
fluctuations (normalized so that $D_+(0)=1$) and 
$\delta_0^{\rm DM}({\bf k})$ is the Fourier transformed DM
linear overdensity at $z=0$. The low-pass 
filter $W_{\rm IGM}(k,z) = (1 + k^2/k_J^2)^{-1}$ depends on 
the comoving Jeans length
\be 
k_J^{-1}(z) \equiv  H_0^{-1} \left[ {2 \gamma k_B T_0(z) \over 3 \mu m_p
\Omega_{0m} (1 + z)}\right]^{1/2} \;,
\ee
with $k_B$ the Boltzmann constant, $T_0$ the temperature at mean
density, $\mu$ the molecular weight of the IGM, $\Omega_{0m}$ the
present-day matter density parameter and $\gamma$ the ratio of
specific heats.  Gnedin \& Hui (1998) adopt a different and more
accurate expression for the IGM filter $W_{\rm IGM}(k,z)$, which,
however, does not allow a simple matching with the non-linear
regime. More accurate window-function have also been proposed  by
Nusser (2000) and Matarrese \& Mohayaee (2001).  In what follows we
take $T_0(z) \propto 1+z$, which leads to a constant comoving Jeans
scale. This assumption should not be critical as the redshift
intervals considered here are small.

Given the simple relation between the IGM and DM linear density
contrasts, one gets the following relation between the corresponding 
linear power spectra, 
$P_0^{\rm IGM}(k,z) = D_+^2(z) W_{\rm IGM}^2(k,z) P(k)$, 
where  $P(k)$ is the  DM power spectrum linearly extrapolated to 
$z=0$. \\

To enter the non-linear regime, BD97 adopt a simple lognormal (LN) model
(Coles \& Jones 1991) for the IGM local density,  
\be
n_{\rm IGM}({\bf x},z) = {\overline n}_{\rm IGM}(z)
\left( 1 + \delta^{\rm IGM}({\bf x}, z) \right) = {\overline n}_{\rm IGM}(z)
\exp\left[\delta_0^{\rm IGM}({\bf x}, z) -
{\langle (\delta_0^{\rm IGM})^2 \rangle D_+^2(z) \over 2} \right] \;,
\ee
where ${\overline n}_{\rm IGM}(z) \approx 1.12 \times 10^{-5}
\Omega_{0b} h^2 (1+z)^3$ cm$^{-3}$.

As stressed by BD97, the LN model for the IGM
has two important features: on large scales, $k\ll
k_J$, it reduces to the correct linear evolution, while on strongly
non-linear scales, $k \gg k_J$, it behaves locally like the isothermal
hydrostatic solution for the intra-cluster gas 
(e.g. Bahcall \& Sarazin 1978),  
$n_{\rm IGM} \sim {\overline n}_{\rm IGM} \exp[-(\mu m_p \Phi_0 /
\gamma k_B T_m)]$, where $\Phi_0$ is the linear peculiar gravitational 
potential. 

The IGM peculiar velocity ${\bf v}^{\rm IGM}$ is related to the linear
IGM density contrast via the continuity equation. As in BD97, we 
assume that the peculiar velocity is still linear even on scales 
where the density contrast gets non-linear; this yields
\be
{\bf v}^{\rm IGM} ({\bf x},z) =  E_+(z) \int {d^3k \over (2\pi)^3}
e^{i {\bf k} \cdot {\bf x}} {i {\bf k} \over k^2} W_{\rm IGM}(k,z) 
\delta_0^{\rm DM}({\bf k}) \;, 
\ee
with
$E_+(z) = H(z)f(\Omega_m,\Omega_\Lambda)D_+(z)/(1+z)$. 
Here $f(\Omega_m,\Omega_\Lambda) \equiv - d\ln D_+(z)/d \ln (1+z)$ 
(e.g. Lahav {\it et al.} 1991, for its explicit and general expression)
and $H(z)$ is the Hubble parameter at redshift $z$, 
\be
H(z) = H_0 \sqrt{\Omega_{0m}(1+z)^3 + \Omega_{0{\cal R}}(1+z)^2 +
\Omega_{0\Lambda}}
\ee
where $\Omega_{0\Lambda}$ is the vacuum-energy contribution to the
cosmic density and $\Omega_{0{\cal R}}=1-\Omega_{0m} -
\Omega_{0\Lambda}$ ($\Omega_{0{\cal R}}=0$ for a flat universe).

\subsection{Line of sight random fields}

If we now draw a LOS in the ${\bf x}_\parallel$ direction, with fixed
coordinate ${\bf x}_\perp$ \footnote {In what follows we neglect the
effect of the varying distance between the lines of sight.}, we obtain
a set of one-dimensional random fields, which will be denoted by the
subscript $\parallel$.  Consider, for instance, the IGM linear density
contrast: for a fixed ${\bf x}_\perp$ we can Fourier transform it
w.r.t. the $x_\parallel$ coordinate and  obtain
\be
\delta^{\rm IGM}_{0\parallel}(k_\parallel,z |{\bf x}_\perp) 
= D_+(z) \int {d^2 k_\perp 
\over (2\pi)^2} e^{i {\bf k}_\perp \cdot {\bf x}_\perp} 
W_{\rm IGM}(\sqrt{k_\parallel^2 + k_\perp^2},z) 
\delta_0^{\rm DM}(k_\parallel,{\bf k}_\perp) 
\equiv D_+(z) \Delta^{\rm IGM}(k_\parallel,z|{\bf x}_\perp) \;. 
\ee
Similarly, for the IGM peculiar velocity along the ${\bf x}_\parallel$
direction, we obtain 
\be
v_\parallel^{\rm IGM}(k_\parallel,z |{\bf x}_\perp) =   
{\bf v}^{\rm IGM}(k_\parallel,z|{\bf x}_\perp) \cdot 
{\hat {\bf x}}_\parallel \equiv  
i k_\parallel E_+(z) U^{\rm IGM}
(k_\parallel,z |{\bf x}_\perp) \;,
\ee
where ${\hat {\bf x}}_\parallel$ is the unit vector along the LOS and
\be 
U^{\rm IGM} (k_\parallel,z |{\bf x}_\perp) 
=  \int {d^2 k_\perp \over (2\pi)^2} 
e^{i {\bf k}_\perp \cdot {\bf x}_\perp} {1 \over k_\parallel^2 +
k_\perp^2} W_{\rm IGM}(\sqrt{k_\parallel^2+k_\perp^2},z) 
\delta_0^{\rm DM}(k_\parallel, {\bf k}_\perp) \;. 
\ee 

\subsubsection{Line of sight auto-spectra and cross-spectra}

We now want to obtain auto and cross-spectra for these 1D Gaussian
random fields along single or multiple LOS. 
Given a 3D random field $\psi({\bf x})$ with Fourier transform 
$\psi({\bf k})$ and 3D power spectrum $P(|k|)$, one can
define the LOS random field $\psi_\parallel(x_\parallel,{\bf x}_\perp)$
as the 1D Fourier transform  
\be
\psi_\parallel(k_\parallel| {\bf x}_\perp) \equiv \int {d^2 k_\perp
\over (2\pi)^2} e^{i{\bf k}_\perp \cdot {\bf x}_\perp}
\psi(k_\parallel,{\bf k}_\perp) \;. 
\ee
The cross-spectrum $\pi(|k_\parallel||r_\perp)$ for our LOS random field 
along parallel LOS, separated by a 
transverse distance $r_\perp$, is defined by
\be
\langle \psi_\parallel(k_\parallel|{\bf x}_\perp) 
\psi_\parallel(k^\prime_\parallel|{\bf x}_\perp + 
{\bf r}_\perp) \rangle = 2 \pi \delta_D(k_\parallel 
+ k^\prime_\parallel) \pi(|k_\parallel||r_\perp) \;, 
\ee
where $\delta_{\rm D}$ is the Dirac delta function and 
$\pi(k|r_\perp)$ can be related to the 3D 
power spectrum as follows 
\be
\pi(k|r_\perp) = \int {d^2 k_\perp \over (2\pi)^2} 
e^{i {\bf k}_\perp \cdot {\bf r}_\perp} P(\sqrt{k_\perp^2 + k^2}) \;. 
\label{eq:gencross}
\ee
Integrating over angles and shifting the integration variable yields
\be
\pi(k|r_\perp) = {1 \over 2 \pi} 
\int_k^\infty dq q J_0(r_\perp\sqrt{q^2 - k^2}) P(q) \;,
\label{eq:cross}
\ee 
where $J_n$ will generally denote the Bessel function of order $n$. 

In the limit of vanishing distance between the two LOS $J_0 \to 1$ and
the above formula reduces to the standard relation for the LOS (1D)
auto-spectrum in terms of the 3D power spectrum (Lumsden {\it et al.}
1989) 
\be 
p(k) \equiv \pi(k|r_\perp = 0) = {1 \over 2\pi}
\int_k^\infty dq q P(q) \;.
\label{eq:autosp}
\ee

The IGM linear density contrast and peculiar velocity along each 
LOS can be grouped together in a single Gaussian random vector field 
${\bf V}(k_\parallel|{\bf x}_\perp)$ with  components  
$V_1 \equiv \Delta^{IGM}$ and $V_2 \equiv U^{IGM}$. 
One has
\be
\langle V_i(k_\parallel|{\bf x}_\perp) V_j(k_\parallel^\prime|{\bf
x}_\perp + {\bf r}_\perp) \rangle
= 2 \pi ~\delta_D (k_\parallel + k_\parallel^\prime)
\pi_{ij}(|k_\parallel||r_\perp) \;, \ \ \ \ \ \ \ \ \ \ \ \ \ \ \ i,j=1,2 \;,
\ee
where the $2 \times 2$ symmetric cross-spectra matrix
$\pi_{ij}(k|r_\perp)$ has components 
\be
\pi_{11}(k|r_\perp) = {1 \over 2\pi} 
\int_k^\infty dq q J_0(r_\perp\sqrt{q^2 - k^2}) W_{\rm IGM}^2(q,z) P(q) \;,
\label{crosssp}
\ee 
\be
\pi_{22}(k|r_\perp) = {1 \over 2\pi} 
\int_k^\infty {dq \over q^3} J_0(r_\perp\sqrt{q^2 - k^2}) 
W_{\rm IGM}^2(q,z) P(q) \;,
\ee 
\be
\pi_{12}(k|r_\perp) = {1 \over 2\pi} 
\int_k^\infty {dq \over q} J_0(r_\perp\sqrt{q^2 - k^2}) 
W_{\rm IGM}^2(q,z) P(q) \;. 
\ee
For vanishing transverse distance between the lines, $J_0 (0) =1$ 
and the auto-spectra components are given by
$p_{ij}(k) \equiv \pi_{ij}(k|r_\perp=0)$, which we will also need in 
the following.    

\subsection{Correlation procedure}

Our next problem is how to generate the two random 
fields $\Delta^{\rm IGM}$ and $U^{\rm IGM}$ in 1D 
Fourier space ($-\infty < k_\parallel < \infty$). These  
random fields have non-vanishing  cross-correlations but 
unlike in 3D Fourier space they cannot be related by simple 
algebraic transformations.  

We gan generally write any  M-dimensional
Gaussian random vector ${\bf V}$ with 
correlation matrix ${\bf C}$ and components   
$c_{ij} = \langle V_i V_j \rangle$, 
as a linear combination of another  
M-dimensional Gaussian random vector ${\bf X}$ 
with diagonal  correlation matrix\footnote{
Simple applications of this general `correlation procedure' in the
M=2 case are given e.g. in (Bi 1993) and (Porciani {\it et al.}
1998).}, which we can take as the identity 
${\bf I}$ without any loss of generality. 
The transformation involves the M $\times$ M  matrix ${\bf A}$, 
with components $\alpha_{ij}$, as follows:
${\bf V} = {\bf A} {\bf X}$. One  gets
${\bf C} = {\bf A} {\bf A}^T$, i.e. 
$c_{ij}  = \sum_k \alpha_{ik} \alpha_{jk}$. 
There is a slight complication because 
${\bf V}$ is a random vector ${\it field}$ 
defined in 1D Fourier space. We can, however,  extend 
the above formalism to  vector fields, assuming 
that ${\bf X}$ is a Gaussian vector
field with white-noise power spectrum,     
\be 
\langle X_i(k_\parallel)  X_j(k_\parallel^\prime) \rangle = 2 \pi 
~\delta_{ij}~\delta_D(k_\parallel + k_\parallel^\prime) \;
\ee
($\delta_{ij}$ is the Kronecker symbol). 
We then have $V_i = \sum_{j=1}^2 \alpha_{ij} X_j$ 
where 3 of the 4 $\alpha_{ij}$ components are determined by the 
conditions $\sum_{k=1}^2 \alpha_{ik} \alpha_{jk} = p_{ij}$.   
The remaining freedom (due to the symmetry of the original
correlation matrix) can be used to simplify the calculations. 
A simple choice of coefficients which solves our problem is 
\be
\alpha_{11} = \sqrt{p_{11} -p_{12}^2/p_{22}} \;, \ \ 
\alpha_{12} = p_{12}/\sqrt{p_{22}} \;, \ \
\alpha_{21} = 0 \;, \ \ 
\alpha_{22} = \sqrt{p_{22}} \;.  
\ee

\subsection{Multiple lines of sight}
 
It is straightforward to extend our formalism to 
simulate the IGM properties along parallel LOS. 
Let  ${\bf V}(k_\parallel)$ and ${\bf W}(k_\parallel)$ be two 
1D Gaussian random vector fields obtained as in Section 2.3,
each with the same set of coefficients $\alpha_{ij}$  
but starting from two independent white-noise
vector fields ${\bf X}$ and ${\bf Y}$ 
(i.e. such that $\langle X_i Y_j \rangle =0$). 
Then both ${\bf V}$ and ${\bf W}$ have the
correct LOS auto-spectra by construction 
while their mutual cross-spectra vanish: 
$\langle V_i (k_\parallel) W_j(k^\prime_\parallel) \rangle
= 0$. 

Let us further define a new vector 
${\bf V}^\prime(k_\parallel|r_\perp)$ with components 
$V_i^\prime= \sum_{k=1}^2\left( \beta_{ik} V_k + \gamma_{ik} W_k
\right)$,
such that its auto and cross-spectra components are given by, 
\ba \langle V_i(k_\parallel) V_j(k_\parallel^\prime) \rangle & = &
\langle V^\prime_i(k_\parallel|r_\perp) V^\prime_j(k_\parallel^\prime|r_\perp)
\rangle = 2 \pi ~\delta_D (k_\parallel + k_\parallel^\prime)
p_{ij}(|k_\parallel|) \;, \nonumber \\
\langle V_i(k_\parallel) V^\prime_j(k_\parallel^\prime|r_\perp) 
\rangle & = &  
2 \pi ~\delta_D (k_\parallel + k_\parallel^\prime)
\pi_{ij}(|k_\parallel||r_\perp) \;. 
\ea

The vectors ${\bf V}$ and ${\bf V}^\prime$ will then represent
our physical IGM linear fields on parallel LOS at a distance 
$r_\perp$. They will be statistically indistinguishable from 
those obtained by drawing two parallel LOS separated 
by $r_\perp$ in a 3D realization of the linear IGM density
and velocity fields.   

The transformation coefficients are determined by the
equations $\sum_{k,\ell=1}^2 \left(\beta_{ik}\beta_{j\ell} + 
\gamma_{ik}\gamma_{j\ell} \right) p_{k\ell} = p_{ij}$ and \linebreak$\sum_{k=1}^2 \beta_{ik} p_{kj} =  \pi_{ij}$. 
Once again, due to the symmetry of the cross-spectra components, 
we can choose one of the four $\gamma_{ij}$
coefficients arbitrarily. The explicit form of $\beta_{ij}$ and 
chosen set of   $\gamma_{ij}$ is 
\ba
\beta_{11} & = & {\pi_{11} p_{22} - \pi_{12} p_{12} \over p_{11} p_{22}
-p_{12}^2} \;, \ \ \ \ \ \ \ \ \ \ \ \ \ \ \ \
\beta_{12} = {\pi_{12} p_{11} - \pi_{11} p_{12} \over p_{11} p_{22}
-p_{12}^2} \;, \nonumber \\
\beta_{21} & = & {\pi_{21} p_{22} - \pi_{22} p_{12} \over p_{11} p_{22}
-p_{12}^2} \;, \ \ \ \ \ \ \ \ \ \ \ \ \ \ \ \ 
\beta_{22} = {\pi_{22} p_{11} - \pi_{12} p_{12} \over p_{11} p_{22}
-p_{12}^2} \;
\ea
and
\ba
\gamma_{11} & = & \pm \sqrt{{p_{22} \over A_{22}}{(A_{11}A_{22} -A_{12}^2) 
\over (p_{11} p_{22} - p_{12}^2)}} \;, \ \ \ \ \  
\gamma_{12} = {1 \over \sqrt{ p_{22} A_{22}}} 
\left[ A_{12} \mp p_{12} \sqrt{A_{11}A_{22} -A_{12}^2 
\over p_{11} p_{22} - p_{12}^2} \right] \;, \nonumber \\
\gamma_{21} & = & 0 \;, \ \ \ \ \ \ \ \ \ \ \ \ \ \ \ \ \ \
\ \ \ \ \ \ \ \ \ \ \ \ \ \ \ \ \ \ 
\gamma_{22} = \sqrt{A_{22} /p_{22}} \;,
\ea 
where
\be
A_{11} = p_{11} - \left(\beta_{11}^2 p_{11} + 2 \beta_{11} 
\beta_{12} p_{12} + \beta_{12}^2 p_{22} \right) \;,
\ee
\be
A_{12} = p_{12} - \beta_{11} \left(\beta_{21} p_{11} +
\beta_{22} p_{12}\right) - \beta_{12} \left(\beta_{21} p_{21} +
\beta_{22} p_{22} \right)
\;,
\ee
\be
A_{22} = p_{22} - \left(\beta_{21}^2 p_{11} + 2 \beta_{21} 
\beta_{22} p_{12} + \beta_{22}^2 p_{22} \right) \;. 
\ee 

With this technique we can produce large ensembles of spatially 
correlated LOS pairs with both high resolution and large redshift 
extent. In this way we can  fully account for the effects of 
cosmic variance on LOS properties. 

The same technique can  be extended to obtain multiple LOS at the
obvious cost of more and more complicated transformation
coefficients. Alternatively,  a two dimensional array of LOS 
in a region of the sky could be simulated. This will be described 
in a future paper. Here we only consider the case of LOS pairs.

\section{Flux correlations in absorption spectra of QSO pairs}

\subsection{Simulating the flux distribution of the Ly$\alpha$ forest}

To simulate the Ly$\alpha$ forest one needs the local neutral hydrogen
density $n_{\rm HI}({\bf x},z)$ and the corresponding 
Ly$\alpha$ optical depth $\tau(z_0)$. In the optically thin limit, 
assuming photoionization
equilibrium, the local density of neutral hydrogen can be written as a
fraction $f_{\rm HI}(T,J_{21},n_{\rm e})$ of the local hydrogen
density $n_{\rm H}({\bf x},z)$, which in turn is a fraction $X\approx
0.76$ of the total baryon density $n_{\rm IGM}({\bf x},z)$.  Here
$\Gamma_{-12}$, the hydrogen photoionization rate in units of
$10^{-12}$ s$^{-1}$, is defined in terms of the UV photoionizing
background radiation as $4\times J_{21}$, where $J(\nu) = J_{21}
(\nu_0 / \nu)^{m} \times 10^{-21} {\rm erg}~{\rm s}^{-1} {\rm Hz}^{-1}
{\rm cm}^{-2} {\rm sr}^{-1}$ with $\nu_0$, the frequency of the HI
ionization threshold, and $m$ is usually assumed to lie between 1.5
and 1.8; $n_{\rm e}({\bf  x},z)$ is the local number density of free 
electrons.  In the highly ionized case 
($n_{\rm HI} \ll n_{\rm IGM}$) of interest here, one can
approximate the local density of neutral hydrogen as (e.g. Hui, Gnedin
\& Zhang 1997) \be n_{\rm HI}({\bf x}, z) \approx 10^{-5} ~{\overline
n}_{\rm IGM}(z) \left({\Omega_{0b} h^2 \over 0.019}\right)
\left({\Gamma_{-12} \over 0.5}\right)^{-1} \left(T({\bf x},z) \over
10^4 {\rm K} \right)^{-0.7} \left({1+z \over 4}\right)^3 \left(1 +
\delta_{\rm IGM}({\bf x},z) \right)^2 \;.  \ee The temperature of the
low-density IGM is determined by the balance between adiabatic cooling
and photoheating by the UV background, which establishes a local
power-law relation between temperature and density,  $T({\bf
x},z) = T_0(z) (1+\delta^{\rm IGM}({\bf x},z))^{\gamma(z)-1}$, where
both the temperature at mean density $T_0$ and the adiabatic index
$\gamma$ depend on the IGM ionization history (Meiksin 1994; Miralda-Escud\'e \&
Rees 1994; Hui \& Gnedin 1997; Schaye {\it et al.} 2000). The
absorption optical depth in redshift-space at $u$ (in km s$^{-1}$) is
\be \tau(u)={\sigma_{0,\alpha} ~c\over H(z)} \int_{-\infty}^{\infty}
dy\, n_{\rm HI}(y) ~{\cal V}\left[u-y-v_{\parallel}^{\rm
IGM}(y),\,b(y)\right]dy \;, \ee where $\sigma_{0,\alpha} = 4.45 \times
10^{-18}$ cm$^2$ is the hydrogen Ly$\alpha$ cross-section, $H(z)$ is
the Hubble constant at redshift $z$, $y$ is the real-space coordinate
(in km s$^{-1}$), ${\cal V}$ is the standard Voigt profile normalized
in real-space, $b=(2k_BT/mc^2)^{1/2}$ is the velocity dispersion in
units of $c$.  For the low column-density systems considered here 
the Voigt profile is well aproximated by a Gaussian:
${\cal V}=(\sqrt{\pi} b)^{-1}\exp[-(u-y-v_{\parallel}^{\rm
IGM}(y))^2/b^2]$. As stressed by BD97 peculiar velocities  affect
the optical depth in two different ways: the lines are shifted to a
slightly different location and their profiles are altered by
velocity gradients.  The quantity $\Gamma_{-12}$ is treated 
as a free parameter, which is tuned in order to match the observed effective
opacity $\tau_{\rm eff}(z) = -\ln \langle \exp{(-\tau)}\rangle$
(e.g. McDonald {\it et al.} 1999; Efstathiou {\it et al.} 2000) at the
median redshift of the considered range ($\tau_{\rm eff} = 0.12$ and
$\tau_{\rm eff}=0.27$ at $z=2.15$ and $z=3$, respectively, in our
case).  We account for this constraint by averaging over the ensemble
of the simulated LOS.  The transmitted flux is then simply ${\cal
F}=\exp(-\tau)$.  Let us finally mention that Bi (1993)
simulated double LOS with a simplified scheme 
which neglects the effects of peculiar velocities.

\subsection{Absorption spectra of QSO pairs in cold dark matter
models}
\begin{figure*}
\resizebox{1\textwidth}{!}{\includegraphics{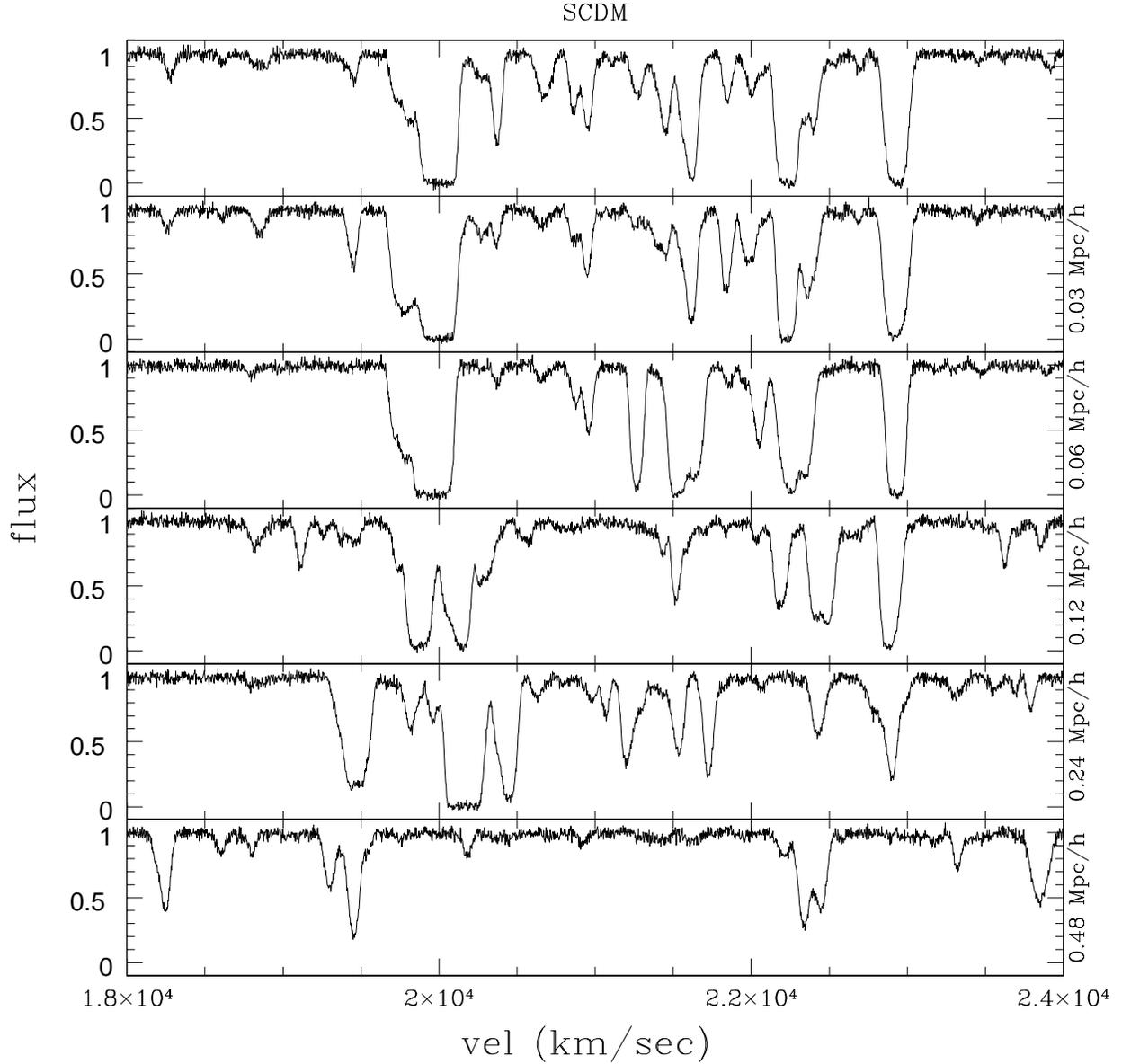}} 
\caption{Simulated QSO spectra for a SCDM model
at $z=2.15$. The y-axis represents the flux,
the x-axis is in km s$^{-1}$. Each of the bottom 
five panels has the correct  correlation properties with
regard to the top panel at proper distance 
$0.03$ Mpc/h, $0.06$ $h^{-1}$ Mpc,
$0.12$ $h^{-1}$ Mpc, $0.24$ $h^{-1}$ Mpc, $0.48$ $h^{-1}$ Mpc, 
from top to bottom.}
\label{fig1}
\end{figure*}

\begin{figure*}
\resizebox{1\textwidth}{!}{\includegraphics{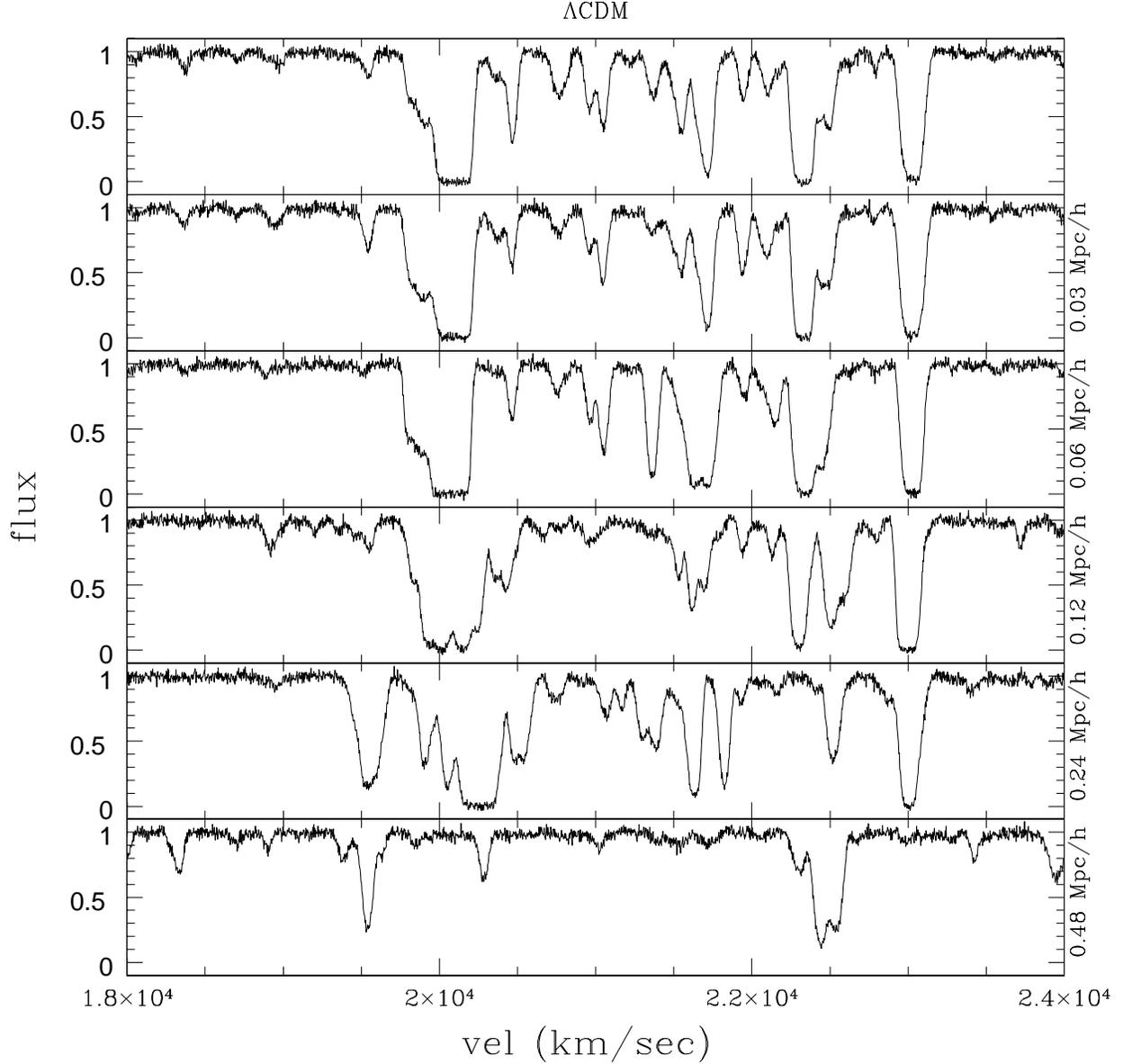}} 
\caption{The same as in Figure \ref{fig1} for a $\Lambda$CDM model, 
with the same random phases.}
\label{fig2}
\end{figure*}

We have simulated a set of LOS pairs all based on the cold dark matter
(CDM) model but with different values of the normalization,
$\sigma_8$, vacuum energy content, $\Omega_{0\Lambda}=1 -
\Omega_{0m}$, Hubble constant $h=H_0/100 {\rm\, km\, s^{-1}\,
Mpc^{-1}}$ and spectral shape-parameter $\Gamma$. A linear
power spectrum of the form $P(k) \propto k\, T^2(k)$ was assumed, with
$T(k)$ the CDM transfer function (Bardeen {\it et al.} 1986): \be
T(q)=\frac{\ln(1+2.34q)}{2.34q}\times\left[1+3.89q+(16.1q)^2+(5.46q)^3
+ (6.71q)^4\right]^{-1/4}\;, \ee where $q=k/h\, \Gamma$. The
shape-parameter $\Gamma$ depends on the Hubble parameter, matter
density $\Omega_{0m}$ and baryon density $\Omega_{0b}$ (Sugiyama
1995): \be \Gamma = \Omega_{0m}\,h
\exp[-\Omega_{0b}-\sqrt{h/0.5}\,\Omega_{0b}/ \Omega_{0m})]\; .  \ee We
have simulated a cluster-normalized Standard CDM model (SCDM)
($\Omega_{0m}=1$, $h=0.5$, $\sigma_8=0.6$, $\Omega_{0b}=0.019\,h^2$),
a $\Lambda$CDM model ($\Omega_{0m}=0.3$, $\Omega_{0\Lambda}=0.7$,
$h=0.65$, $\Omega_{0b}=0.019\,h^2$) and a $\tau$CDM model
($\Omega_{0m}=1$, $h=0.5$, $\sigma_8=0.6$, $\Omega_{0b}=0.019\,h^2$,
$\Gamma=0.187$).

The redshift ranges of the Ly$\alpha$ forest are $1.90\leq z \leq
2.40$ ($3525$ \AA$ < \lambda < 4133 $\AA) and $2.75 \leq z \leq 3.25$
($4556$ \AA $< \lambda < 5163$\AA).  We use 1D grids with
$2^{14}=16,384$ equal comoving-size intervals. In the first case the
box length is 538 comoving Mpc for the SCDM and 718 comoving Mpc for
the $\Lambda$CDM model, while in the second case is 378 comoving Mpc for
the SCDM and 518 comoving Mpc for the $\Lambda$CDM. These intervals
have been chosen so that the size of the box, expressed in km
s$^{-1}$, is the same for all the models in each redshift interval. In
the low-redshift case our box size is 47690 km s$^{-1}$, while in the
high-redshift one is 37540 km s$^{-1}$.

The adopted procedure to account for observational and instrumental
effects follows closely that described in (Theuns, Schaye \& Haehnelt
1999).  We convolve our simulated spectra with a Gaussian with full
width at half maximum of FWHM = 6.6 km s$^{-1}$, to mimic QSO
spectra as observed by the HIRES spectrograph on the Keck telescope.
We then resample each line to pixels of size 2 km s$^{-1}$.  Photon 
and pixels noise is finally added, in such a way that the
signal-to-noise ratio is approximately $50$ but it varies as a
function of wavelength and flux of observed QSO spectra
as estimated from a  spectrum of $Q1107+485$.

Figures \ref{fig1} and \ref{fig2} show the transmitted flux for a 
sequence of LOS with varying transverse distance $r_\perp$, for SCDM
and $\Lambda$CDM model respectively. In each sequence the first LOS 
is kept fixed (the one on the top) and the value of
$r_\perp$ varies in the cross-spectra while the phases are kept 
constant; the second member of each pair has thus the required
auto and cross-correlation properties. Notice that only pairs which
include the first LOS have the required cross-correlation
properties.

The figures are very similar because we use the same phases in both
models, so that the differences can be better appreciated. 
Coherent structures extend out to hundreds of  kpc/h proper
(several comoving Mpc) in the direction orthogonal to the LOS. 
These can be understood as the signature of the underlying `cosmic
web' of mildly non-linear sheets and filaments 
(Bond, Kofman \& Pogosyan 1996) in the dark matter distribution, which
is smoothly traced by low-column density Ly$\alpha$ absorption
systems.

\subsection{Statistical analysis of the flux correlations}

\begin{figure*}
\resizebox{0.8\textwidth}{!}{\includegraphics{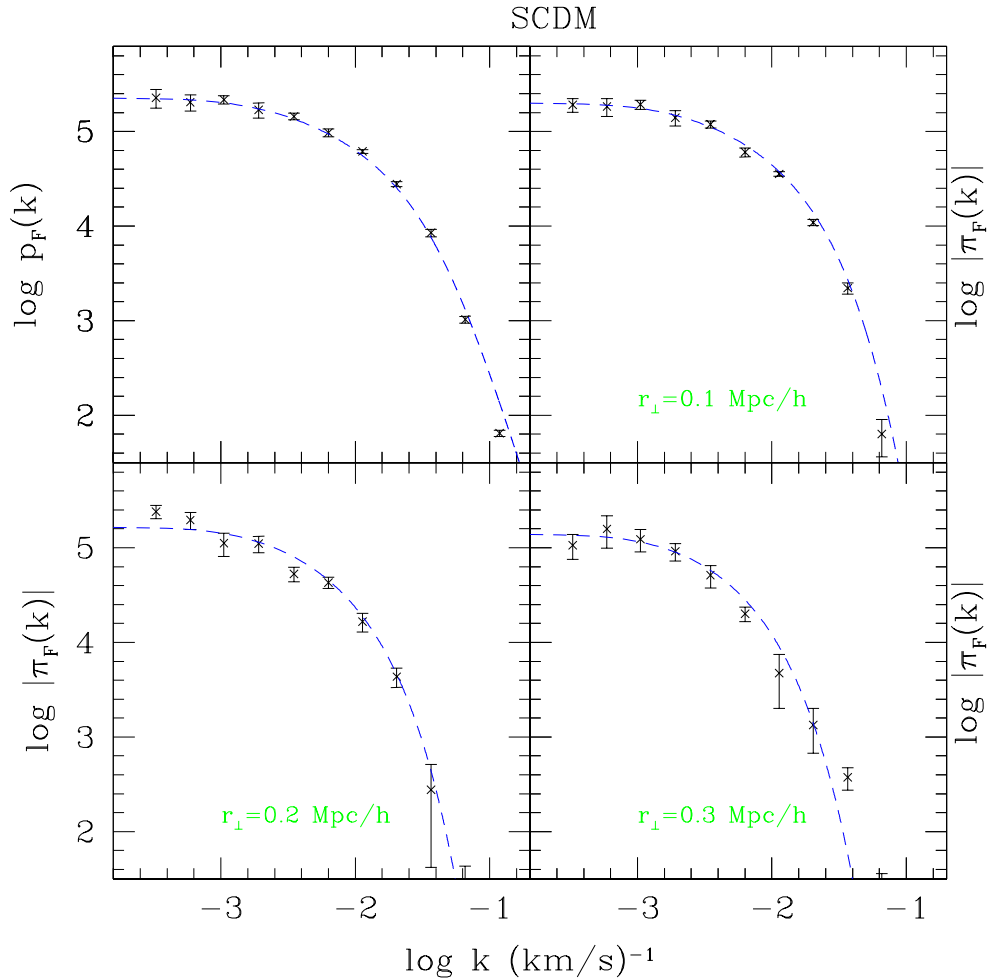}} 
\caption{Auto-spectrum (top panel on the left) and three flux
cross-spectra obtained with the definitions given in the text, for the
SCDM model. The x-axis represents $\log k$, where $k$ is defined as
$2\pi/v$, the y-axis is the log of auto-spectra and cross-spectra. The
three cross-spectra have been obtained taking 10 QSO pairs whith three
different proper separations ($0.1$ $h^{-1}$ Mpc, $0.2$ $h^{-1}$ Mpc,
$0.3$ $h^{-1}$ Mpc). The dashed curve is the linear prediction of the
1D auto-spectrum and cross-spectrum as given in eq. (\ref{eq:autosp})
and in eq.  (\ref{crosssp}). Error bars are the error of the mean
value.}
\label{fig3}
\end{figure*}

\begin{figure*}
\resizebox{0.8\textwidth}{!}{\includegraphics{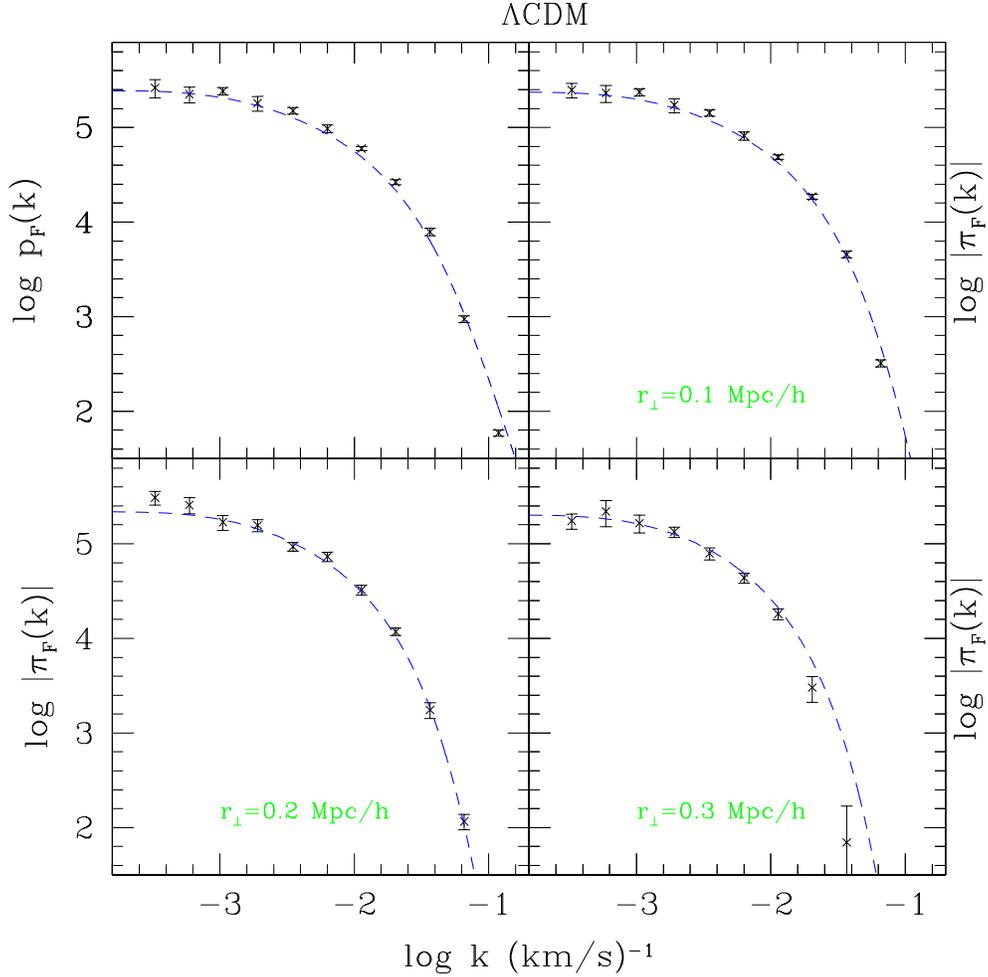}} 
\caption{Same plot as in Figure \ref{fig3} for a
$\Lambda$CDM model}
\label{fig4}
\end{figure*}

We compute the auto-spectra $p_{\cal F}(k)$ and the cross-spectra
$\pi_{\cal F}(k|r_{\perp})$ of the flux for the two cosmological
models using the definitions $p_{\cal F}(k)=\langle|{\cal
F}_0(k)|^2\rangle$ and $\pi_{\cal F}(k|r_{\perp})=\langle Re\, ({\cal
F}_0(k) {\cal F}_1^*(k,r_{\perp})) \rangle$, where ${\cal F}_0(k)$ and
${\cal F}_1(k)$ are the Fourier components of the flux along the two
LOS at distance $r_{\perp}$ and the symbol $\langle \cdot \rangle$
denotes the ensemble average. In Figures \ref{fig3} and \ref{fig4} we
plot the auto-spectrum (top panel on the left) and the three
cross-spectra at proper distance $0.1$ $h^{-1}$ Mpc, $0.2$ $h^{-1}$
Mpc and $0.3$ $h^{-1}$ Mpc.  The total number of simulated LOS pairs
is 30.  The results are ensemble averages of 10 pairs at each
separation and the error bars represent the scatter of the mean value.
The range of $k=2\pi/v$ ($v$ is the velocity in km/s) plotted here
does not include the small scales (high $k$) strongly affected by
pixel noise and non-linearity effects where the power spectra flattens
again (Theuns, Haehnelt \& Schaye 2000, McDonald {\it et al.}
2000). The dashed line represents the theoretical prediction of the
linear power spectra as given by equation ($\ref{crosssp}$). The
agreement is good over a wide range of wavenumbers $k$, roughly $-3\,
\mincir \log k \, \mincir -1$. This is the interval we will use in
Section 6 to recover the 3D power spectrum of the linear density field
and is close to the range of k-wavenumbers used by Croft {\it et al.}
(1999) from the analysis of observational data. 

It is important here to stress that the simulated spectra have been
produced in redshift-space, while the theory is in real-space. We have
checked the difference by recomputing eq. (\ref{eq:autosp}) and
eq. (\ref{crosssp}) considering also redshift-space distortions,
i.e. using the distortion kernel proposed by Hui (1999), and the
differences are negligible. Given this reasonably good agreement, all
the following comparisons between the simulated spectra and the theory
have been made without taking into account the redshift-space
distortions in the theoretical equations.
 
We have also measured the flux cross-correlation coefficient
$\chi(r_\perp)$ as a function of separation, binning 
the data in bins with width $\Delta v$. We choose 
the following definition,  
\be
\chi(r_\perp)=\frac{1}{N_{pix}}\frac{\sum_{i=1}^{N_{pix}} ({\cal
F}_0(i)-<{\cal F}_0>)({\cal F}_1(i)-<{\cal F}_1>)}{\sigma_{{\cal F}_0}
\sigma_{{\cal F}_1}}, 
\ee 
where ${\cal F}_0$ and ${\cal F}_1$ are the
fluxes of the binned spectra of pairs 
with  separation $r_{\perp}$, $N_{pix}$ is the number
of pixels of the binned spectrum and $\sigma_{{\cal F}_0}$,
$\sigma_{{\cal F}_1}$ the standard deviation of the two fluxes.

This function can be related to the auto and cross-spectra as follows,
\be
\chi(r_\perp) = {\int_0^\infty dk ~e^{-k^2/k_{\rm
s}^2} \pi(k|r_\perp) \over  \int_0^\infty dk  ~e^{-k^2/k_{\rm s}^2}
p(k)} \;,
\label{ccc_def}
\ee where $k_{\rm s}^{-1} \propto (\Delta v)^{-1}$.  
Note that the flux cross-correlation coefficient
does not depend on the amplitude but only on the shape 
of the power spectra. Using this function we can define a transverse 
coherence scale $r_{{\rm c}\perp}$ as the distance between two LOS at 
which $\chi(r_{{\rm c}\perp}) = 0.5$. Analytical estimates of 
$\chi(r_\perp)$ for the various CDM models can be obtained 
by replacing  $p$ and $\pi$ with the IGM linear density auto and 
cross-spectra $p_{11}$ and $\pi_{11}$ in the above relation.

We compare here three cosmological models, a SCDM, a $\Lambda$CDM and
a $\tau$CDM model ($\Gamma \sim 0.2$) at seven    angular distances 
(15, 30, 40, 55, 65, 80, 100 arcsec) at redshift $2.15$.
This corresponds to different comoving distances $r_{\perp}$ 
in the three different cosmologies (e.g. Liske 2000),   
\be 
r_\perp =
\frac{c\,\theta}{H_0}\int_0^{z}\left[\Omega_{0m}(1+z')^3 +
\Omega_{0\Lambda}\right]^{-1/2}\,d\,z' \;\;\;\;\; 
\label{angdist}
\ee 
The IGM temperature at mean density
and the temperature-density relation parameter  
are assumed to be $T_0=10^{4.2}$ K and $\gamma=1.3$, 
respectively.

\begin{figure*}
\resizebox{1\textwidth}{!}{\includegraphics{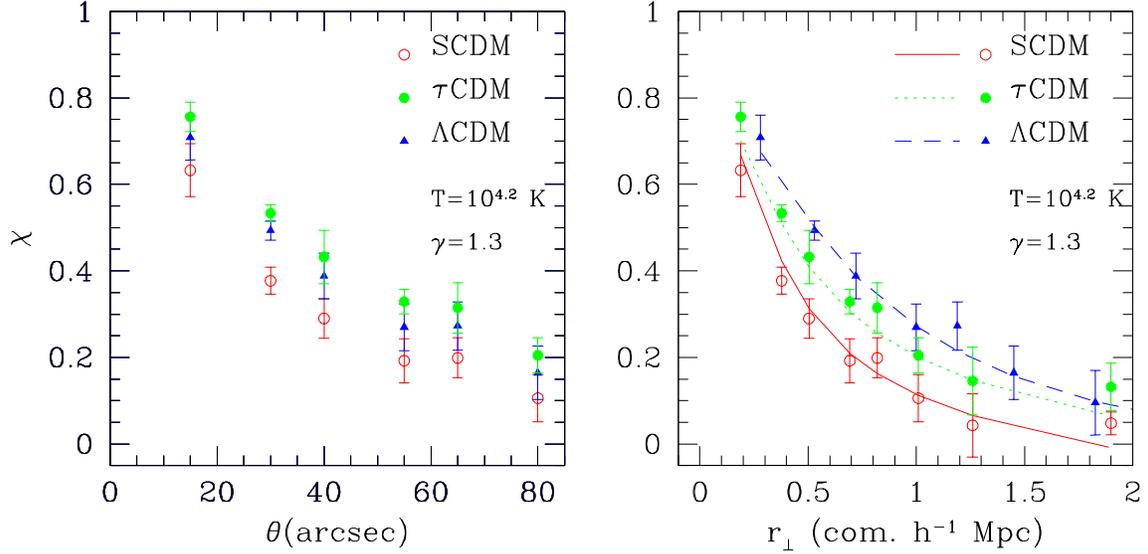}}
\caption{Cross-correlation coefficient $\chi$ plotted as a function
of angular separation $\theta$ (left panel) and comoving separation  
$r_{\perp}$ (right panel). 
Points are the outputs of simulated QSO
pairs for a SCDM (open circles) a $\tau$CDM with $\Gamma\sim 0.2$ (filled
circles) and a $\Lambda$CDM model (triangles). The results are
averaged over 8 pairs for each  $\theta$  (15, 30,
40, 55, 65, 80). The  curves in the right panel are the theoretical
predictions of equation $(\ref{ccc_def})$. The dotted curve is for
$\tau$CDM, the dashed is for $\Lambda$CDM while the solid curve is
for SCDM. The error bars represent the scatter of the distribution.}
\label{fig5}
\end{figure*}

We have generated 8 pairs of spectra for each angular separation in the
usual way. The results are plotted in Figure \ref{fig5} 
for the three models. The left and right panels show the 
cross-correlation coefficient  against 
angular and comoving separation, respectively.  
The cross-correlation coefficient has been
calculated directly from the whole spectrum
(binned with $\Delta v \sim 3$ km s$^{-1}$).
The three solid curves are calculated using  
equation $(\ref{ccc_def})$, where 
$\pi(k|r_{\perp})$ and $p(k)$ are replaced with the corresponding
quantities for the IGM ($\pi_{11}$ and $p_{11}$).  The agreement with
the theoretical prediction (with $k_s \sim 2\pi/3$ km$^{-1}$ s) is
reasonably  good. 

Using  $\chi=0.5$ as the definition for the coherence length 
gives  $0.10 \pm 0.04 $ $h^{-1}$ Mpc for
SCDM, $0.13 \pm 0.03$ $h^{-1}$ Mpc for $\tau$CDM and $0.17 \pm 0.05$
$h^{-1}$ Mpc for $\Lambda$CDM (proper).  
The cross-correlation coefficient depends on the detailed shape of the
IGM power spectrum at and above the Jeans length.  In the SCDM model
the power spectrum falls most steeply towards larger scales and this
results in the shortest coherence length of the three models.  The
$\tau$CDM model with its flatter power spectrum on the relevant scales
has a significatly larger coherence length. The $\Lambda$CDM model has
an even larger coherence length due to the larger Jeans length for
smaller $\Omega_{0m}$ at fixed temperature (eq. 2). However, the plot
which can be directly compared with observations is the one on the
left in Figure \ref{fig5}, which does not invoke any {\it a priori}
assumption on the cosmological model.  Unfortunately, the
$\Omega_{0m}$ dependence becomes negligible if the cross-correlation
coefficient is plotted against angular separation.  The
cross-correlation cofficient can, however, be used to constrain the
shape of the DM power spectrum (which we have chosen here to
paramerize with $\Gamma$) if the temperature $T_0$ and the temperature
density relation coefficient $\gamma$ are determined indepedently.

We have also  run simulations with a wider range of model 
parameters, changing $\Gamma$, $\Omega_{0m}$, $\sigma_8$ and 
the parameters describing the physics of the IGM, such as $T_0$, 
$\gamma$, $\Gamma_{-12}$.  If $\tau_{eff}$ is fixed over the whole 
ensemble of simulations the dependencies on the amplitude of
the power spectrum at that redshift and on $\Gamma_{-12}$
largely cancel. 

The coherence length as defined above 
does  not rely on assumptions about the shape 
of the absorbers.   If the IGM indeed traces the filaments 
in the underlying dark matter distribution this should be a
more adequate measure  of the `characteristic size' of  the absorbers
than the usually performed coincidence analysis of absorption lines  
fitted with a Voigt profile routine (see next section).

\section{Coincidence analysis of Ly$\alpha$ absorption lines in QSO pairs}

In this section we perform a coincidence analysis of absorption lines 
in the spectra along adjacent lines of sight 
(see Charlton {\it et al.} 1997 for a corresponding  analysis 
of a hydrodynamical simulation). We have used the Voigt profile 
fitting routine AUTOVP (Dav\'e et al. 1997) to identify
and characterize the absorption lines.  For this analysis 
we have generated absorption spectra of  5 QSO pairs in the range 
$1.9 \le z \le 2.4$ for proper distances $0.05, 0.1, 0.2, 0.4$ 
and $0.6$ $h^{-1}$ Mpc  in the $\Lambda$CDM model only. 

Charlton et al. (1997) demonstrated that for the filamentary and 
sheet-like absorping structures expected in hierachical 
structure formation scenarios the characteristic absorber 
`size', as determined  by counting  `coincident' and 
`anticoincident' lines in QSO pairs, will depend on 
the separation of the QSO pair and on the column density 
threshold used. The characteristic size determined in this 
way is thus of little physical  meaning and difficult to interpret.  
Nevertheless, such an analysis is useful in order to make connections
with published observational studies which usually 
perform such an analysis.

\begin{figure*}
\resizebox{1\textwidth}{!}{\includegraphics{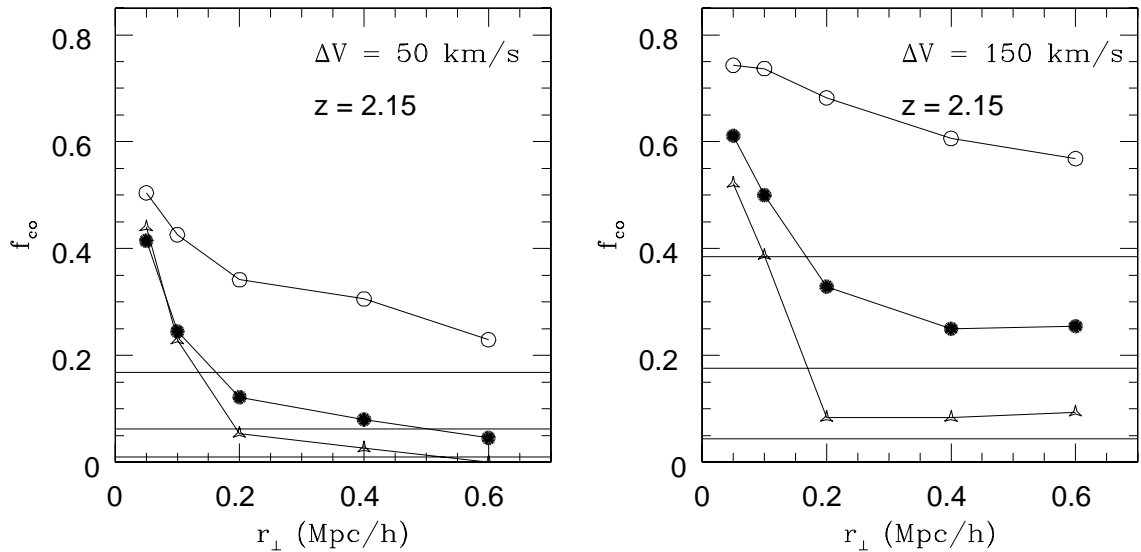}}
\caption{The ratio of the number of coincidences to the number of
coincidences plus anticoincidences, $f_{co}$, as a function of the proper
separation of QSOs in  $\Lambda$CDM model at a median redshift
$z=2.15$.  Left and right panel are for a velocity difference 
 $\Delta v=50$ km/s and $\Delta v=150$ km/s, respectively. 
The three solid curves show $f_{co}$ for
lines with a column density higher than $N_{\rm HI} = 10^{12}$
cm$^{-2}$, $10^{13}$ cm$^{-2}$ and $10^{14}$ cm$^{-2}$ (from top to
bottom).  The three solid horizontal lines represent
the level of  random coincidences as estimated from $10$ pairs
of uncorrelated LOS, for the same column density thresholds.}
\label{hm1}
\end{figure*}

\begin{figure*}
\resizebox{1\textwidth}{!}{\includegraphics{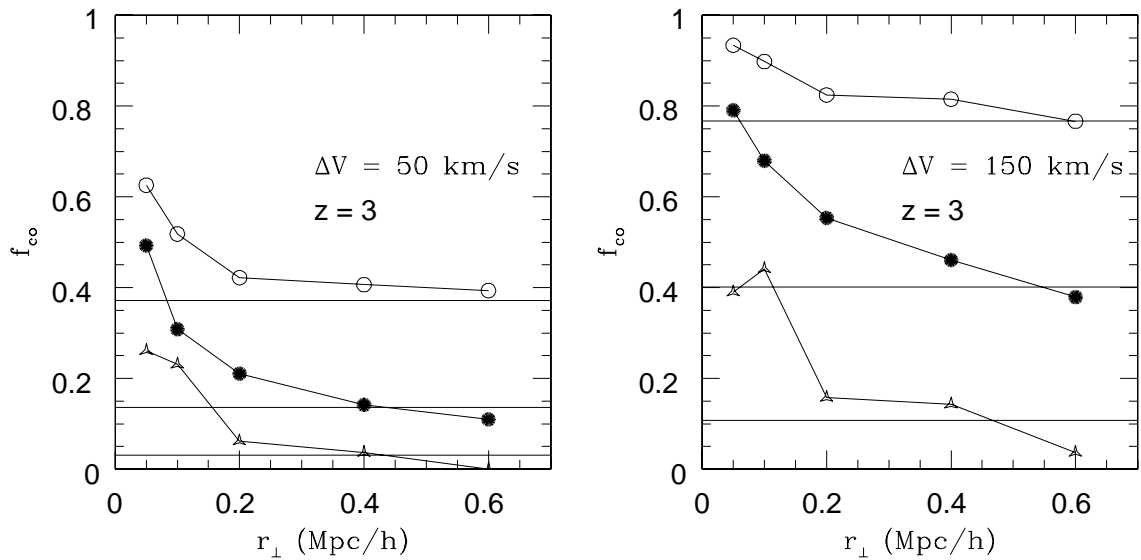}}
\caption{The same as in Figure \ref{hm1} but at redshift $z=3$.}
\label{hm1b}
\end{figure*}

We adopt the  definition of Charlton {\it et al.} (1997) 
based on the  `hits-and-misses' statistics
described in McGill (1990). A coincidence is defined as the case in
which an absorption line is present in both spectra within a given 
velocity difference $\Delta v$ and above some signal-to-noise ratio. 
An anticoincidence is defined 
when a line is present in one but not in the other spectrum. 
If there are two lines within $\Delta v$ we count only one 
coincidence and no anti-coincidence, as in Fang (1996). 
We also generate 5 uncorrelated LOS and compute the number 
of coincidences and anticoincidences in these 10 pairs to estimate 
the level of random  coincidences.

In Figure \ref{hm1} we plot the quantity $f_{co}$, the ratio of 
the number of coincidences to the  sum of all  coincidences and 
anticoincidences as a function of proper distance. The three 
curves are  for  different column density thresholds 
($10^{12}$, $10^{13}$ and $10^{14}$ cm$^{-2}$) and the left 
and right panels are for $\Delta v = 50,\, 150$ km s$^{-1}$, 
respectively.  The solid horizontal lines show the level of random 
coincidences as estimated from the ensemble average of 
$10$ uncorrelated spectra. The curves depend strongly on the 
choice of the velocity  difference $\Delta v$.

In Figure \ref{hm1b} we plot the same quantities but for spectra 
with a median redshift $z=3$, i.e. the LOS span the range
$2.75 < z < 3.25$. At fixed redshift, $f_{co}$ is larger if the 
 velocity difference $\Delta v$ allowed for a coincidence is 
larger. This is easily understood as the chance to get a
`hit' becomes higher. There is also a significant trend with redshift.
With increasing redshift both $f_{co}$ and the level of 
random coincidences increase, the latter by a factor of two. 
These findings are similar to those of  Charlton {\it
et al.}  (1997) (their Figure 2).  Our  level of  
random coincidences is somewhat smaller than that 
in Charlton {\it et al.} (1997), probably due to a different temperature
which results in a larger Jeans scale. 
\begin{figure*}
\resizebox{0.8\textwidth}{!}{\includegraphics{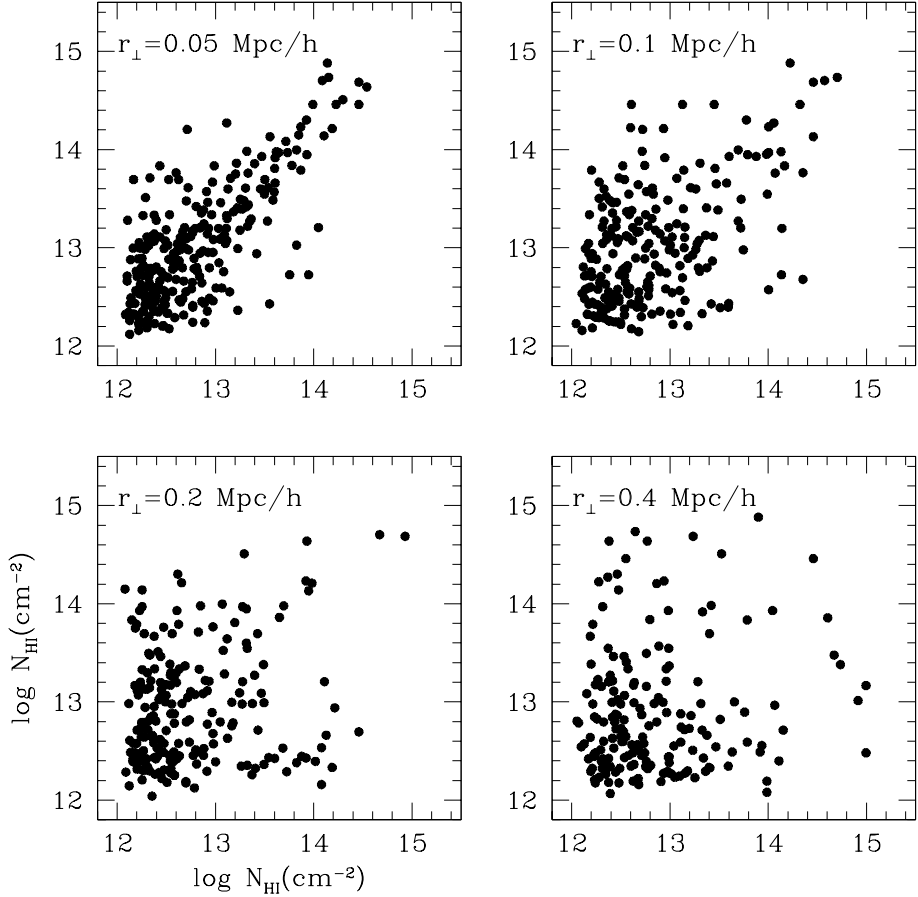}}
\caption{Column density of neutral hydrogen, $\log N_{\rm HI}$, for
pairs of coincident absorption lines identified with 
the Voigt profile  fitting routine AUTOVP
in  a $\Lambda$CDM model
at $z=2.15$.  A velocity difference of $\Delta v= 50$ km/s 
was allowed for coincident lines. 
Four different proper separations are shown:
$r_{\perp}=0.05$ $h^{-1}$ Mpc, top left panel; $r_{\perp}=0.1$ $h^{-1}$ Mpc, top
right panel; $r_{\perp}=0.2$ $h^{-1}$ Mpc, bottom left panel and
$r_{\perp}=0.4$ $h^{-1}$ Mpc, bottom right panel.}
\label{hm3}
\end{figure*}

In  Figure \ref{hm3} we show scatter plots of the neutral 
hydrogen column densities  of coincident lines 
above a column density threshold of $10^{12} \rm cm^{-2}$. 
For the smaller separations the column densities are well correlated 
while for larger distances the column density difference rapidly 
increases. This is again not surprising as for large separations 
the `coincidences' occur mostly by chance. 
The same analysis has been done for lines found at $z=3$.
The  result is very similar to the one found at $z=2.15$.

The results presented in this section depend on details 
of the Voigt profile fitting and  the velocity difference and 
column density threshold chosen to do the `hits-and-misses' 
statistics. This makes it difficult  to infer physically 
meaningful properties of the absorbers, as for example their 
characteristic size, with these techniques. The coherence 
length defined in the last section is more useful in this 
respect. 

\section{Recovering the 3D dark matter power spectrum using 
flux cross-spectra} 

\subsection{A new method for obtaining the 3D dark matter spectrum from
the flux auto-spectrum}    

If the effect of peculiar velocities, thermal broadening and
instrumental noise on the flux fluctuations at small scales are
neglected the transmitted flux at redshift $z$ in a given direction
${\hat \theta}$ can be approximated as (e.g. Croft {\it et al.} 1998;
Theuns {\it et al.} 1999) 
\be {\cal F}({\hat \theta}, z) =
\exp\left[-A \left(1 + \delta^{\rm IGM} \left({\bf x}({\hat
\theta},z), z\right)\right)^\beta\right] \;, 
\label{FGPA}
\ee 

where $\beta \approx 2 - 0.7(\gamma-1)$, while $A$ is a normalization
constant of order unity which determines the mean flux in the
considered redshift interval. Eq. (\ref{FGPA}) is valid only in
redshift-space and the effect of peculiar velocity and thermal
broadening increases the scatter in the relation between ${\cal F}$ and
$\delta^{\rm IGM}$ (Croft {\it et al.} 1998).

If smoothed on a sufficiently 
large scale the IGM overdensity can  be treated 
as a linearly fluctuating field. In this case 
the fluctuations of the flux,  $\delta{\cal F}$, are simply 
related to  the linear baryon density perturbations,  
\be 
\delta {\cal F} ({\hat \theta}, z) \approx - A \beta  
\delta_0^{\rm IGM}\left({\bf x}({\hat \theta}, z),z\right) \;. 
\ee

\begin{center}
\begin{figure*}
\resizebox{1\textwidth}{!}{\includegraphics{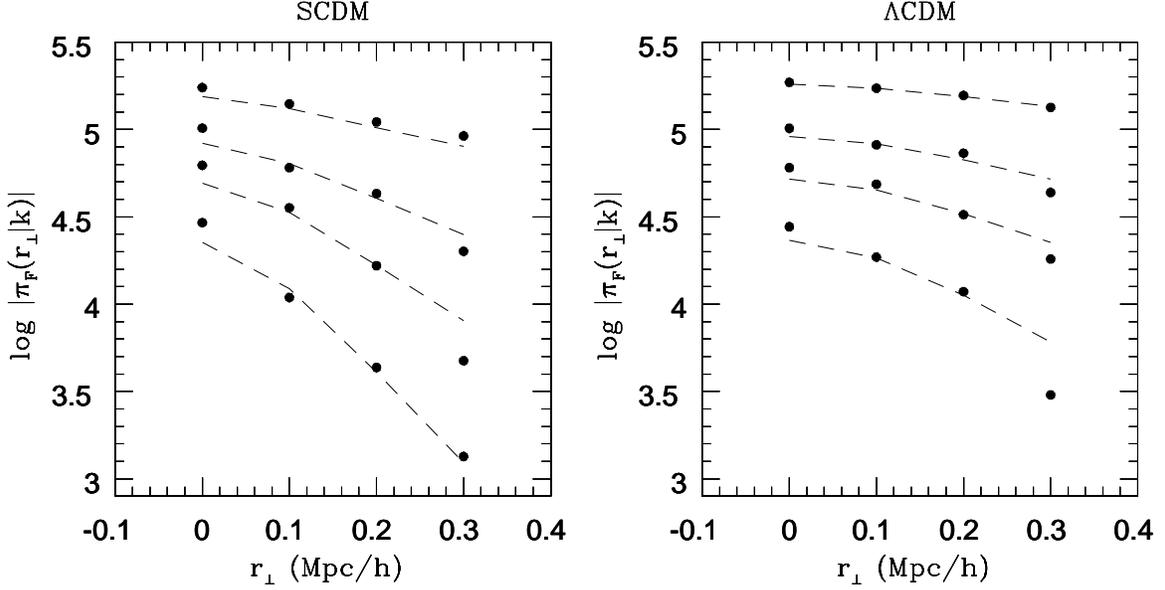}} 
\caption{$\pi_F$ as a function of $r_{\perp}$ (in proper $h^{-1}$ Mpc
at $z=2.15$) for 4 different values of $k$: $k=0.169, 0.561, 1.005,
1.806$ Mpc$^{-1}$ for SCDM model (left panel, from top to bottom) and
$k=0.125, 0.414, 0.742, 1.333$ Mpc$^{-1}$ for the $\Lambda$CDM model
(right panel, from top to bottom).  The dashed curve is the
theoretical prediction given by eq. (\ref{crosssp}), not corrected for
redshift-space distortions. The points are calculated for an ensemble
of 30 realizations, 10 realizations for each separation: $r_{\perp}=0$
$h^{-1}$ Mpc (which gives the auto-spectrum) and $r_{\perp} = 0.1,
0.2, 0.3$ $h^{-1}$ Mpc.}
\label{fig_pi}
\end{figure*}
\end{center}

On large scales  observed absorption spectra can thus be used 
to recover the 3D primordial power spectrum of DM perturbations. 
The standard procedure suggested  by Croft 
{\it et al.} (1998, 1999, 2000) inverts the relation in eq. (\ref{eq:autosp}) 
to  obtain  the 
3D power spectrum by {\it differentiating} the 1D auto-spectrum,   
\be
P(k) = - {2 \pi \over k} {d \over dk} p(k) \;. 
\label{eq:recdiff} 
\ee 
Alternative methods to measure the amplitude 
of DM fluctuations and their power spectrum have been proposed 
by Hui (1999) and by Nusser \& Haehnelt (1999, 2000).

However, the 3D power spectrum can also be reconstructed by {\it
integrating} the 1D cross-spectrum $\pi(k |r_\perp)$ over the
transverse separation between LOS pairs. Indeed, by inverse Fourier
transforming eq. (\ref{eq:gencross}) on the plane spanned by ${\bf
r}_\perp$, and integrating over angles, we  find
\be
P(k) = 2 \pi \int_0^\infty d r_\perp r_\perp J_0(r_\perp\sqrt{k^2 - q^2})
~\pi(q |r_\perp) \equiv \int_0^\infty {d r_\perp \over r_\perp} 
~{\cal Q}(r_\perp|k,q) \;, \ \ \ \ \ \ \ \ q\leq k \;. 
\label{eq:recgen}
\ee
The above results also lead to a useful `consistency relation' between
the LOS auto-spectrum $p(k)$ and the cross-spectra $\pi(k|r_\perp)$
along LOS pairs,  
\be 
\pi(k|r_\perp) = p(k) - r_\perp
\int_0^\infty dq J_1(qr_\perp) p\left(\sqrt{q^2 + k^2}\right) \;.
\label{check}
\ee

As a consequence of the assumed homogeneity and isotropy, 
the RHS of equation (\ref{eq:recgen}) does not depend on $q$;   
we can therefore simplify the integral  
by taking  $q=k$,  
\be
P(k) = 2 \pi \int_0^\infty d r_\perp r_\perp \pi(k |r_\perp) \;.
\label{eq:recint}
\ee

The flux cross-spectrum $\pi_F(k|r_{\perp})$ is plotted in Figure
\ref{fig_pi} as a function of $r_{\perp}$ for a SCDM (left panel) and
a $\Lambda$CDM (right panel) model, from an ensemble of 30 pairs at
different separations. The different set of points are for different
values of $k$ in the range for which the agreement with theoretical
predictions is good. The $r_{\perp}=0$ $h^{-1}$ Mpc point which
represents the auto-spectrum is also shown. The 3D power spectrum can
be obtained using eq. (\ref{eq:recint}).  To estimate $P(k)$ we can
use the values of $\pi_F(k|r_{\perp})$ obtained from the simulations
in the analytical expression. Also in this case the reconstruction
procedure is based on eqs. which neglect the effects of redshift-space
distortions, while the spectra have been produced in redshift
space.

More generally the redundancy shown by the $q$-dependence of the 
integrand in the RHS of eq. (\ref{eq:recgen}) can 
be  exploited to choose a weighting such 
that the dominant contribution to the integral comes 
from the separation range where the signal-to-noise ratio and/or the 
number of observed pairs is highest. 

\begin{figure*}
\resizebox{0.8\textwidth}{!}{\includegraphics{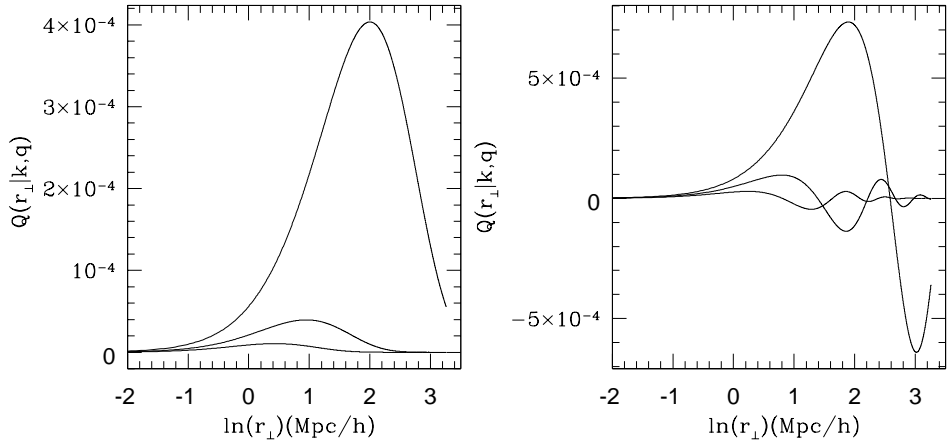}} 
\caption{$Q$ as a function of ln(r$_{\perp}$) in $\Lambda$CDM model for
three values of $k$: $k=0.1251, 0.4142, 0.7422$ Mpc$^{-1}$ (from 
top to bottom) and for two choices of $q$: $q=k$ in the
left panel and $q=k/3$ in the right panel.}
\label{q1}
\end{figure*}

This point is made more clear by Figure \ref{q1}, where
the quantity ${\cal Q}$ is plotted as a function of the transverse
separation $r_\perp$ for various values of $k$ and $q$. In Figure
\ref{q1} the solid curves show the theoretical predictions
for ${\cal Q}$ in a $\Lambda$CDM Universe: in the left panel for $q=k$; 
in the right panel for $q=k/3$. It is evident that the behaviour 
is very different: for $q=k$ the Bessel function ${\rm
J_0}$ is equal to 1, while for $q=k/3$ it acts as a filter which
takes both positive and negative values.

To simplify matters we dropped the explicit redshift dependence of the
power spectra in the above discussion and used power spectra which
were averaged over some redshift interval.  The recovered 3D power
spectrum will then generally depend on the median redshift of that
interval.  Moreover, the assumed isotropy is generally broken,
e.g. due to the effect of peculiar velocities.  A more refined
treatment should thus account for various sources of anisotropy in the
reconstruction procedure by applying for example the techniques
proposed by Hui (1999), McDonald \& Miralda-Escud\'e (1999) and Hui
{\it et al.} (1999).  These and other aspects of the power spectrum
reconstruction will be discussed elsewhere.

A reconstruction procedure for the DM power spectrum based on eqs. 
(\ref{eq:recgen}) or (\ref{eq:recint})  has an obvious 
advantage over the standard method based on eq. (\ref{eq:recdiff}): 
one integrates rather than differentiates  a set of generally 
noisy data. To investigate this we have  performed the 
reconstruction of the initial 3D power spectrum
in three different ways: 
\begin{itemize}
\item {\it differentiation} of simulated 1D auto-spectra, based on eq. (\ref{eq:recdiff});
\item {\it integration} of 1D cross-spectra, recovered from simulated
1D auto-spectra, based on eqs. (\ref{check}) and (\ref{eq:recint});
\item {\it fitting} of simulated 1D cross-spectra, based on
eq. (\ref{eq:cross}).
\end{itemize}

\begin{center}
\begin{figure*}
\resizebox{1.\textwidth}{!}{\includegraphics{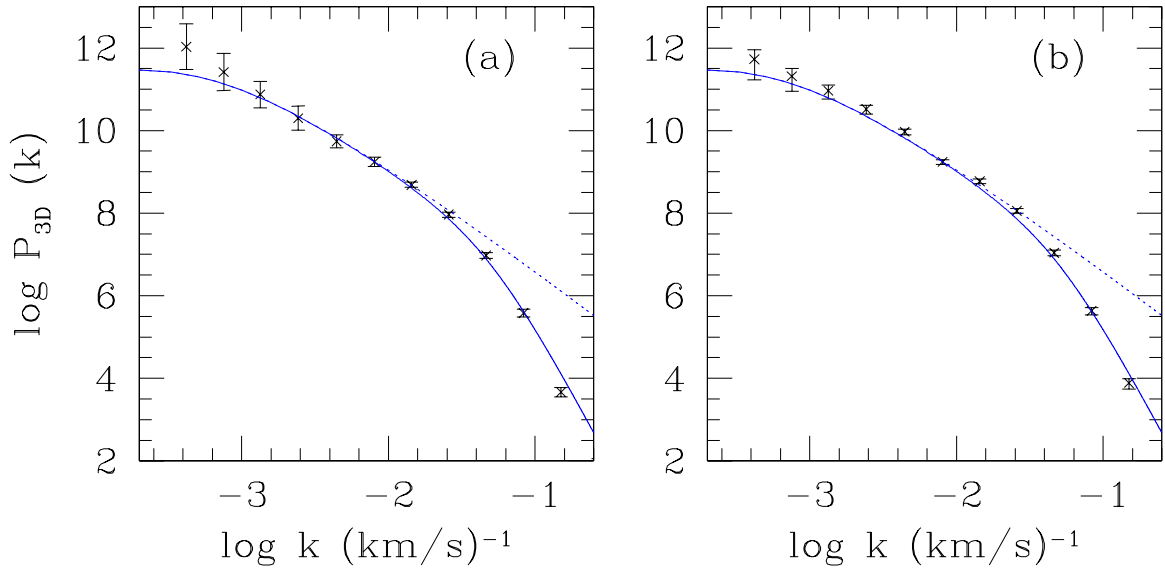}} 
\caption{3D power spectrum reconstruction. In panel $(a)$ points 
represent the 3D power spectrum obtained with `differentiation', 
i.e. using eq. (\ref{eq:recdiff}). 
In  panel $(b)$ points represent the same quantity obtained with
`integration', i. e. using eqs. (\ref{check}) and (\ref{eq:recint}). 
The dotted curve is the 3D power spectrum of the dark matter, 
while the solid curve is the 3D power spectrum of the IGM.} 
\label{diff3637}
\end{figure*}
\end{center}

At first we generate 15 LOS and then compute the flux auto-spectra.
We choose to smooth the auto-spectra with a polynomial function before
using eq. (\ref{eq:recdiff}) to recover the
3D power spectrum. In panel $(a)$ of Figure \ref{diff3637} one can see
that the agreement with the theoretical prediction is good over a wide range of
wave-numbers. The error bars represent the scatter over the
distribution of the 15 recovered 3D power spectra.  Almost all the
$1\sigma$ error bars of the points match the continuous line which
represents the 3D linear IGM power spectrum.  We stress the fact that
this technique has been used `directly' on simulated fluxes, without
the use of further assumptions, such as Gaussianization (e.g. Croft
{\it et al.} 1998).

In panel $(b)$ of Figure \ref{diff3637} we report the results obtained
with the second technique based on equations (\ref{check}) and
(\ref{eq:recint}), using only the auto-spectra information. For each
of the 15 flux auto-spectra we compute the cross-spectra for a large
number of separations via eq. (\ref{check}). Next we use these
estimates in eq. (\ref{eq:recint}), i.e. we integrate the 1D flux
cross-spectra along the transverse direction. The agreement with
linear theory of this `integration' technique is very good. In this
panel the error bars represent the scatter over the distribution of
recovered 3D power spectra.  The results obtained with these two
techniques are basically equivalent provided we smooth the simulated
data.  In a sense this method provides a natural choice of a smoothing
function. If we apply these two methods directly without any smoothing
then `differentiation' is less accurate in recovering the 3D dark
matter power spectrum at large scales and `integration' produces
smaller error bars.  Another significant result is that both methods
fail to recover the 3D dark matter power spectrum for large
wave-numbers ($k\, \magcir 0.1 $ km$^{-1}$ s): at small scales,
peculiar velocities, thermal broadening and nonlinear gravitational
effects are responsible for a drop of the flux power spectrum below that
predicted by  linear theory.

Our third method  uses eq. (\ref{eq:cross}): we fit the
simulated cross-spectra with a $\Lambda$CDM power spectrum with the
amplitude as a free parameter and determine the best estimate of 
the amplitude  by $\chi^2$ minimization.  We thereby fitted 
only the part at small wave-numbers (scales larger than
few Mpc) less affected by  non-linear effects, peculiar
velocities and Jeans length smoothing.  A further advantage of 
using the cross-spectra is that in a set of 
N auto-spectra, there are $N \times (N-1)/2$ cross-spectra, so 
this second sample could in principle be significantly larger 
than the first.  We proceed as follows.
First we calculate the error in estimating the amplitude of the
3D power spectrum by using only 5 flux auto-spectra. Then, having
generated a set of QSO pairs with given separations, we determine the
minimum number of QSO pairs needed to recover the amplitude with the 
same accuracy and with a mean value compatible within $1\sigma$ with
the amplitude recovered from the 5 flux auto-spectra.  We find that
from a set of 5 simulated flux auto-spectra it is possible to
determine the logarithm of the amplitude with an accuracy of $1\%$.
We find that at large separations one needs more pairs to give an
accurate estimate of the amplitude. For separations smaller than 2
comoving Mpc we need less than 10 QSO pairs to reach the same accuracy
and the same amplitude of the auto-spectra estimate.
A detailed method that recovers not only the amplitude but also the
slope of the 3D power spectrum, based on eq. (\ref{eq:cross}) 
will be described in a future paper.

We conclude from our comparison that with the  `integration' technique
the correct slope of 3D dark matter power spectrum can be inferred on 
scales larger than 1 comoving Mpc. This method is complementary to the
usual `differentiation' and is very accurate.  Our `fitting' technique 
for the flux cross-spectra of a  set of QSO pairs can also constrain 
the 3D power spectrum.

\subsection {Cross-spectra as a means of overcoming 
limitations of auto-spectra due to continuum fitting}

As discussed  by Croft {\it et al.}  (1998, 2000)
and Hui {\it et al.} (2000),  at scales $k< 0.005$ km$^{-1}$ s 
corresponding to about $10 - 15\, h^{-1}$ Mpc  the errors in   observed 3D 
flux auto-spectra as determined by a bootstrap 
analysis  increase rapidly. Errors in the continuum 
fitting procedure are likely be the main contributor.  
In principle such continuum fitting errors 
can both increase or decrease  the amplitude of the flux auto-spectrum.

Cross spectra should in principle not be affected by any modulation 
of the flux which is uncorrelated between adjacent LOS 
even though the shot noise will increase with increasing 
fluctuation amplitude.    
Continuum fitting should thus affect flux cross-spectra 
much less than flux auto-spectra. 

In this subsection we verify that this is indeed the case and assess
how many quasar pairs are needed to extend measurements of the DM
spectrum to scales as large as $60\, h^{-1}$ Mpc or more.  We first mimic the
effect of errors in the continuum fitting  procedure on flux auto
and cross-spectra and then go on to demonstrate explicitly that
continuum fitting may actually not be necessary for the analysis of
flux cross-spectra.

To mimic the effect of continuum fitting errors we do the following. 
We take the continuum of the QSO $Q1422+231$  kindly provided by Michael 
Rauch (Rauch {\it et al.} 1997). 
From this we generate a series of spectra 
with continuum fitting errors by changing the amplitude of the 
Fourier modes of our analytical spectra at all k corresponding 
to scales larger  than 15 comoving Mpc, i.e. the scales most 
likely affected by  errors in the continuum fitting procedure. 
We add to the  old amplitude a quantity which varies randomly 
between $-$ 10 \% to 10 \% of the amplitude of 
the corresponding Fourier mode of the continuum of $Q1422+231$. 
Then we calculate  a new spectrum keeping all the phases of the 
Fourier modes of these  continua in order to preserve  
the characteristic emission lines of the continuum of $Q1422+231$. 
We also shift the  spectrum randomly in redshift 
to avoid correlations due to the charateristic emission 
lines. Similarly we produce spectra with different independent 
continua by randomly varying the large scale 
Fourier modes of  the continuum of $Q1422+231$ by $\pm$ -15 \%. 
In the following we will refer to the first set of spectra as 
the {\it no continuum} case and to the second as the {\it continuum} 
case.  We have simulated spectra of 30 QSO pairs in a 
$\Lambda$CDM Universe with an angular 
separation of 100 arcsec in this way.

\begin{center}
\begin{figure*}
\resizebox{0.8\textwidth}{!}{\includegraphics{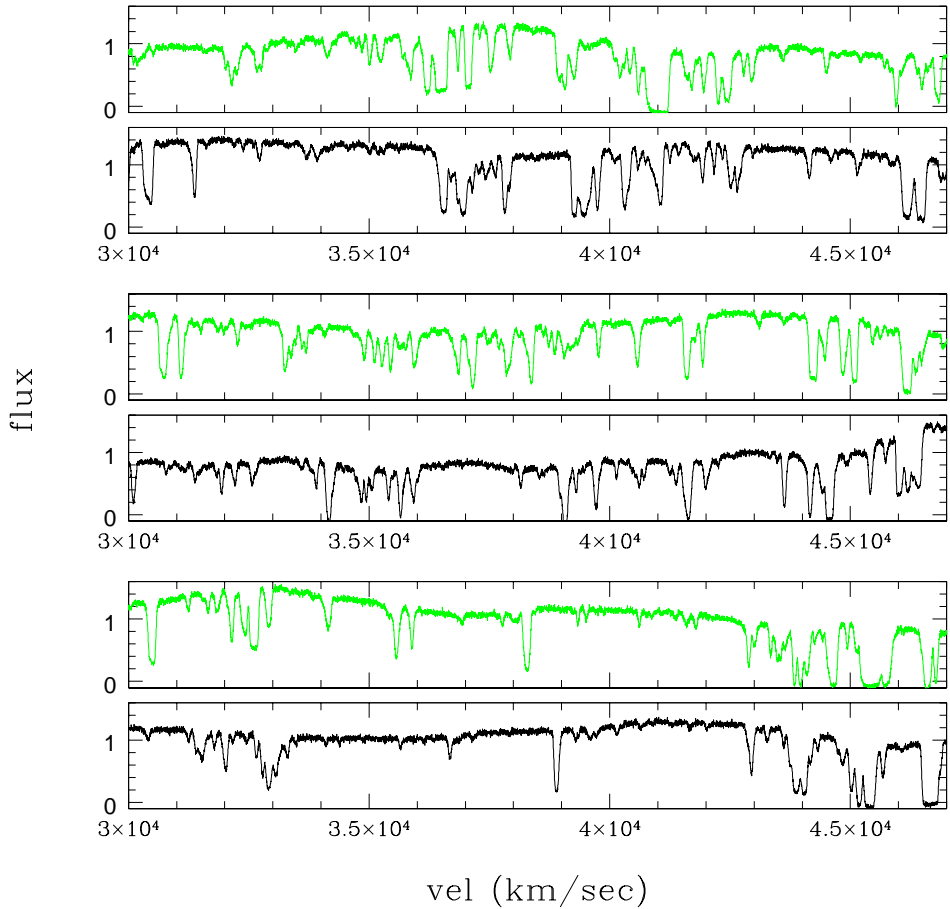}} 
\caption{Spectra from three QSO pairs with continuum errors added as
described in the text}
\label{crpaper}
\end{figure*}
\end{center}

Examples of spectra of three QSO pairs with `continuum fitting errors'  
are shown in Figure \ref{crpaper} (no continuum case). 
Our `continuum fitting errors' are clearly visible by eye 
and are significantly larger  than those that should be achievable  
with a careful continuum fitting procedure. 

\begin{center}
\begin{figure*}
\resizebox{1\textwidth}{!}{\includegraphics{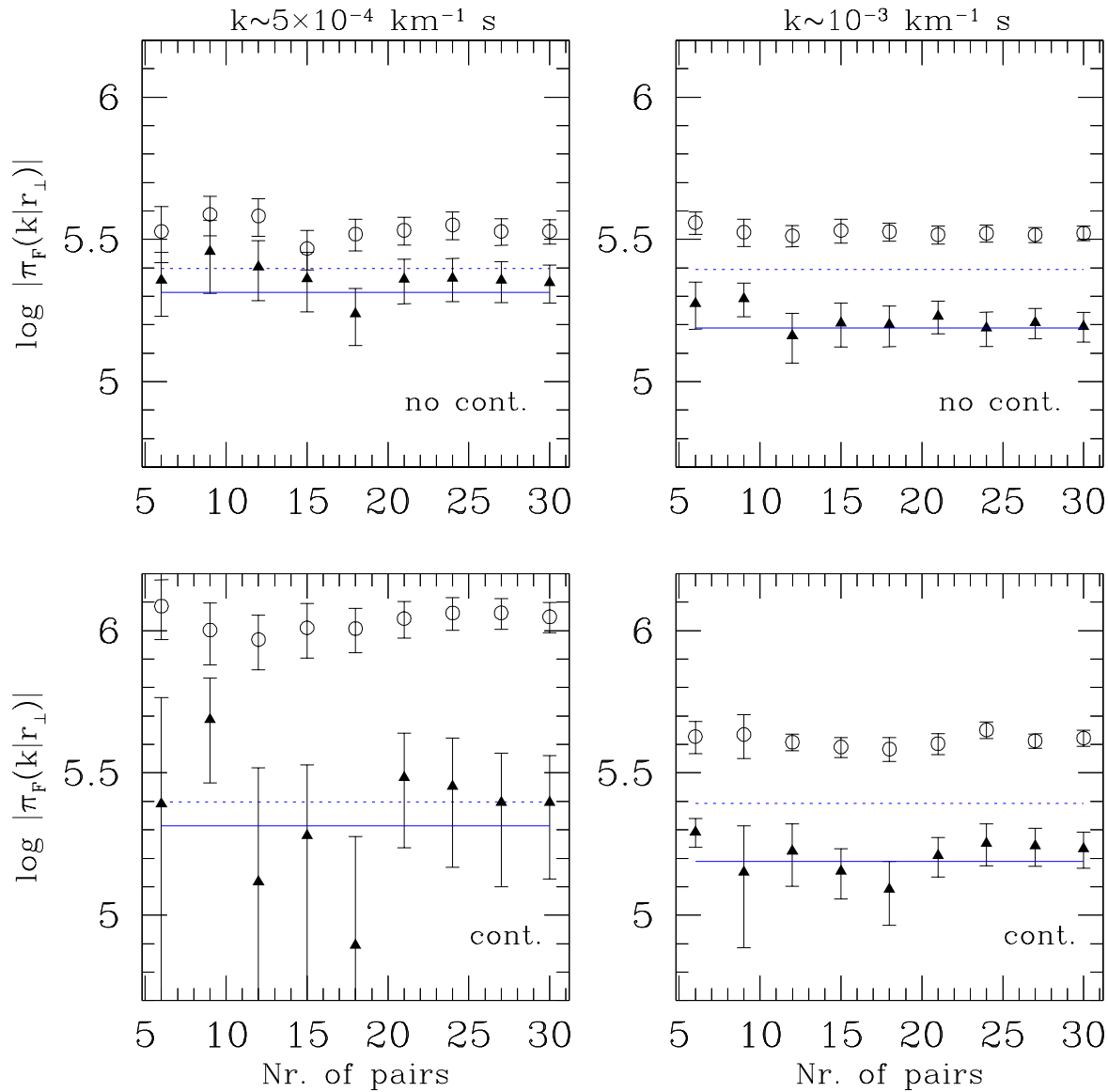}} 
\caption{Flux cross-spectra (triangles) vs. number of QSO
pairs. Points represent the flux auto-spectra. The solid curve is
the theoretical prediction for the IGM linear density cross-spectrum,
eq.  (\ref{crosssp}), while the dotted curve is the theoretical
prediction for the IGM linear density auto-spectrum,
eq. (\ref{eq:autosp}). The left panels are for  $k=5
\times 10^{-4}$ km$^{-1}$ s, corresponding to a scale of about $60\,
h^{-1}$ Mpc; the right panels are for  $k=10^{-3}$ 
km$^{-1}$ s, corresponding to a scale of about 120 $h^{-1}$ Mpc. The
upper panels refer to the {\it no continuum} case, in which no
continuum is present and we mimic the errors introduced by the
continuum-fitting. The bottom panels refer to the {\it continuum}
case, in which we directly cross-correlate spectra without removing
the continuum. Error bars are the error of the mean values. }
\label{crpaper2}
\end{figure*}
\end{center}

We have computed  flux cross-spectra and auto-spectra for the two 
samples. The results are reported in Figure
\ref{crpaper2}, where we  plot the value of the flux
cross-spectra and auto-spectra computed from a given number of pairs,
at two fixed values of $k=5\times10^{-4}$ km$^{-1}$ s (left panels)
and $k=10^{-3}$ km$^{-1}$ s (right panels). These correspond to scales of
$60$ and $120\, h^{-1}$ Mpc, respectively.  The upper panels refer to the
{\it no continuum} case and the bottom ones to {\it continuum} one.

In all four cases shown here the flux auto-spectra (points)
give values significantly larger than those of the linear
prediction (dotted line), as determined by eq. (\ref{eq:autosp}).  
The flux cross-spectra (triangles), however,  converge to the
right mean value (continuous line) obtained from eq. (\ref{crosssp}),
although the number of pairs needed is not small.  This is 
due to the fact that the fluctuations in the continua of the 
QSO eventually cancel out when the spectra are cross-correlated, 
while they add in quadrature to the flux auto-spectrum.

The variations of the flux cross-spectrum around the mean value 
are due to the combined effect of cosmic variance and 
the shot noise due to the continuum fitting errors and the 
continuum fluctuations, respectively. 
At $k=10^{-3}$ km$^{-1}$ s the  1D flux cross-spectrum 
can be obtained  with an accuracy of  2\% in 
logarithmic units from 30 QSO pairs  in both the  
{\it continuum} and the {\it no continuum} case. 
It should thus be possible to constrain the 
rms fluctuation amplitude  of the matter density 
at scales of $60\, h^{-1}$ Mpc  with an accuracy of about 
15\%. Such a measurement should not require 
continuum fitting. At even larger scale  the errors 
for the {\it continuum} case start to blow up with 
30 QSO pairs. A larger number of pairs would be 
required to beat down the shot noise. 

Attempts to recover the 3D power spectrum on large 
scales using the flux auto-spectra clearly do not 
determine the correct value if continuum 
fitting errors are present. As demonstrated above it should, however,  
be possible to  overcome this limitation 
by the use of flux cross-spectra.  

A simple method to recover the right power spectrum for small 
wave-numbers could be  based on eq. (\ref{eq:cross}). We
can parametrize the power spectrum and then compute 
the cross-spectra with $r_{\perp}$  equal to 
the separation of the pairs. The slope of the power spectrum at large 
scales can then be constrained by  $\chi^2$ minimization. A more detailed
discussion on the use of 1D cross-spectra for the determination of the 
DM power spectrum on all scales will be presented elsewhere.

\section{Discussion and conclusions}
 
We have presented  an effective implementation of analytical
calculations of the Ly$\alpha$ opacity distribution of the
Intergalactic Medium (IGM) along multiple lines of sight (LOS) to 
distant quasars in a cosmological setting. We thereby assumed that the
neutral hydrogen distribution traces the dark matter distribution on
scales larger than the Jeans length of a warm photoionized IGM. 
Simulated absorption spectra with varying transverse separation 
between different LOS have been investigated. 
We have  identified coincident 
absorption features as fitted with a Voigt profile fitting routine
and calculated the cross-correlation coefficient and 
the cross-power spectrum of the flux distribution along different LOS   
to quantify the flux correlation.

As expected the correlation of the flux along adjacent 
LOS is sensitive to the detailed shape of the power spectrum 
on scales at and above the Jeans 
length. We have studied the dependence of the cross-correlation
coefficient on the shape parameter $\Gamma$ of the assumed CDM model and
on the Jeans scale which determines the small-scale cut-off of the
power spectrum of the gas distribution due to pressure of the gas. 
We have confirmed previous results that the characteristic size 
of the absorbers inferred from simple hit-and-miss statistics of 
fitted absorption lines assuming spherical 
absorbers depends strongly on the column density threshold used and on the
separation of the QSO pairs. This reiterates the point 
that the filamenatary and sheetlike distribution of the IGM suggested
by numerical simulations makes the concept of spherical absorbers with
a characteristic size of very limited use. Nevertheless, we obtain values  
which are in reasonable agreement with those derived from observations
of multiple systems by Crotts \& Fang (1998), Petitjean {\it et al.} (1998), 
D'Odorico {\it et al.} (1998), Young {\it et al.} (2000).

The cross-correlation coefficient can be used to define a
`characteristic' correlation length of the absorbers in a more
objective way. We obtain $0.10\pm 0.04 $ $h^{-1}$ Mpc for SCDM, $0.13
\pm 0.03$ $h^{-1}$ Mpc for $\tau$CDM and $0.17 \pm 0.05$ $h^{-1}$ Mpc for
$\Lambda$CDM (all at $z=2.15$) as the scale where the cross-correlation
coefficient falls to 0.5. This is about the Jeans length of the 
IGM at this redshift.  We demonstrate that if the temperature and the 
slope of the temperature density relation can be determined 
indepedently then the cross-correlation coefficient can be
used to constrain the shape parameter $\Gamma$ in a way which is 
independent of the amplitude of the power spectrum.

We furthermore propose a new technique to recover the 3D
linear dark matter power spectrum by integrating over 1D flux
cross-spectra.  This method is complementary to the usual
`differentiation' of 1D auto-spectra and suffers different
systematic errors. It can  be used  for the calculation 
of the cross-power spectrum from a given set of auto-power spectra. 
We show that it is mathematically equivalent to the usual 
`differentiation' but offers a natural way of smoothing the data 
in the presence of noise. 

The biggest advantage of the cross-correlation of the flux
distribution of adjacent lines is its ability to eliminate errors
which are uncorrelated in different LOS. The erroneous flux
fluctuations introduced by the continuum fitting procedure, which is
necessary to remove the non-trivial wave-length dependence of the
quasar emission, are such an error which is largely uncorrelated. We
demonstrate that, as expected, such uncorrelated errors affect
the cross-power spectrum significantly less than 
the auto-power spectrum. This may render the tedious and somewhat
arbitrary continuum fitting procedure unnecessary for the recovery of
the dark matter power spectrum from the flux correlations in adjacent
LOS.

Continuum fitting errors have been the main limitation of using flux
auto-power spectra to constrain the DM power spectrum at scales larger
than about $10h^{-1}$Mpc. When flux correlations of adjacent LOS are
used the errors at large scales will be dominated by cosmic variance,
residuals in the removal of the effect of peculiar velocities, the
uncertainty in the temperature density relation, and possible
temperature fluctuations of the IGM which result in opacity
fluctuations due to the temperature dependence of the recombination
coefficient. The errors due to cosmic variance and peculiar velocities
will decrease with increasing number of LOS and flux correlations
should thus allow to extend studies of the DM power spectrum with the
Ly$\alpha$ forest to significant larger scale than is possible with
flux auto-power spectra. We estimate that 30 pairs with separation of
1-2 arcmin are necessary to determine the 1D  cross-spectrum at scales
of $60\, h^{-1}$ Mpc, with an accuracy of about 30\%  if the error 
is dominated by cosmic variance.

\section*{Acknowledgments.} 
We acknowledge Simon White and Saleem Zaroubi for useful discussions.
SM thanks Cristiano Porciani for helpful discussions on the
correlation procedure.  We thank Romeel Dav\'e for making AUTOVP
available. MV acknowledges partial financial support from an EARA Marie
Curie Fellowship under contract HPMT-CT-2000-00132. This work was
supported by the European Community Research and Training Nework `The
Physics of the Intergalactic Medium'.


\begin{thebibliography}{}

\bibitem[]{} Alcock C., Paczy\'nski B., 1979, Nature, 281, 358
\bibitem[]{} Bahcall J.N., Salpeter E.E., 1965, ApJ, 142, 1677 
\bibitem[]{} Bahcall J.N., Sarazin C.L., 1978, ApJ, 219, 781
\bibitem[]{} Bardeen J.M., Bond J.R., Kaiser N., Szalay A.S., 1986, ApJ, 304, 15
\bibitem[]{} Bechtold J., Crotts A.P.S., Duncan R.C., Fang Y., 1994,
ApJ, 437, L83
\bibitem[]{} Bi H.G., 1993, ApJ, 405, 479
\bibitem[]{} Bi H.G., B\"orner G., Chu Y., 1992, A\&A, 266, 1
\bibitem[]{} Bi H.G., Davidsen A.F., 1997, ApJ, 479, 523 
\bibitem[]{} Bi H.G., Ge J., Fang L.-Z., 1995, ApJ, 452, 90 
\bibitem[]{} Bond J.R., Kofman L., Pogosyan D., 1996, Nature, 380, 603
\bibitem[]{} Cen R., Miralda-Escud\'e J., Ostriker J.P., Rauch M.,
1994, ApJ, 437, L83 
\bibitem[]{} Charlton J.C., Anninos P., Zhang Y., Norman M.L., 1997, 
ApJ, 485, 26
\bibitem[]{} Coles P., Jones B., 1991, MNRAS, 248, 1
\bibitem[]{} Croft R.A.C., Weinberg D.H., Katz N., Hernquist L., 1998,
ApJ, 495, 44
\bibitem[]{} Croft R.A.C., Weinberg D.H., Pettini M., Hernquist L., 
Katz N., 1999, ApJ, 520, 1
\bibitem[]{} Croft R.A.C., Weinberg D.H., Bolte M., Burles S., Hernquist L.,
Katz N., Kirkman D., Tytler D., 2000, ApJ, submitted, astro-ph/0012324 
\bibitem[]{} Crotts A.P.S., Fang Y., 1998, ApJ, 502, 16
\bibitem[]{} Dav\'e R., Hernquist L., Weinberg D.H. , Katz N., 1997, 
ApJ, 477, 21
\bibitem[]{} Dinshaw N., Foltz C.B., Impey C.D., Weymann R.J., 
Morris S.L., 1995, Nature, 373, 223 
\bibitem[]{} Dinshaw N., Impey C.D., Foltz C.B., Weymann R.J.,
Chafee F.H., 1994, ApJ, 437, L87
\bibitem[]{} Dinshaw N.,  Weymann R.J., Impey C.D., Foltz C.B., 
Morris S.L., Ake T., 1997, ApJ, 491, 45
\bibitem[]{} D'Odorico V., Cristiani S., D'Odorico S., Fontana A.,
Giallongo E. Shaver P., 1998, A\&A, 339, 678 
\bibitem[]{} Efstathiou G., Schaye J., Theuns T., 2000, 
Philosophical Transactions of the Royal Society, Series A, Vol. 358,
no. 1772,\linebreak p. 2049  
\bibitem[]{} Fang Y., Duncan R.C., Crotts A.P.S., Bechtold J., 
1996, ApJ, 462, 77
\bibitem[]{} Feng L.-L., Fang L.-Z., 2000, preprint astro-ph/0001348
\bibitem[]{} Foltz C.B., Weymann R.J., R\"oser H.-J., Chaffee F.H.,
1984, ApJ, 281, L1
\bibitem[]{} Gnedin N.Y., Hui L., 1996, ApJ, 472, L73  
\bibitem[]{} Gnedin N.Y., Hui L., 1998, MNRAS, 296, 44 
\bibitem[]{} Gunn J.E., Peterson B.A., 1965, ApJ, 142, 1633
\bibitem[]{} Hernquist L., Katz N., Weinberg D.H., Miralda-Escud\'e
J., 1996, ApJ, 457, L51
\bibitem[]{} Hui L., 1999, ApJ, 516, 525
\bibitem[]{} Hui L., Gnedin N.Y., Zhang Y., 1997, ApJ, 486, 599
\bibitem[]{} Hui L., Stebbins A., Burles S., 1999, ApJ, 511, L5
\bibitem[]{} Hui L., Burles S., Seljak U., Rutledge R. E., Magnier E.,
Tytler D., 2000, preprint astro-ph/0005049
\bibitem[]{} Kim T.-S., Cristiani S., D'Odorico S., 2001, preprint 
astro-ph/0101005
\bibitem[]{} Lahav O., Lilje P.B., Primack J.R., Rees M.J., 1991
MNRAS, 251, 128 
\bibitem[]{} Liske J., 2000, MNRAS, 319, 557
\bibitem[]{} Liske J., Webb J.K., Williger G.M., Fern\'andez-Soto A., 
Carswell R.F., 2000, MNRAS, 311, 657
\bibitem[]{} Lumsden, S. L., Heavens, A. F., Peacock, J. A., 1989,
MNRAS, 238, 293
\bibitem[]{} Matarrese S., Mohayaee, R., 2001, preprint
astro-ph/0102220
\bibitem[]{} McDonald P., Miralda-Escud\'e J., 1999, ApJ, 518, 24
\bibitem[]{} McDonald P., Miralda-Escud\'e J., Rauch M., 
Sargent W.L.W., Barlow A., Cen R., Ostriker J.P., 2000, ApJ, 543, 1
\bibitem[]{} McGill C., 1990, MNRAS, 242, 544
\bibitem[]{} Meiksin A., 1994, ApJ, 431, 109
\bibitem[]{} Miralda-Escud\'e J., Rees M., 1994, MNRAS, 266. 343 
\bibitem[]{} Miralda-Escud\'e J., Cen R., Ostriker J.P., Rauch M.,
1996, ApJ, 471, 582
\bibitem[]{} Narayanan V.K., Spergel D.N., Dav\'e R., Ma C.P., 2000,
preprint astro-ph/0001247
\bibitem[]{} Nusser A., 2000, MNRAS, 317, 902
\bibitem[]{} Nusser A., \& Haehnelt M., 1999, MNRAS, 303, 179
\bibitem[]{} Nusser A., \& Haehnelt M., 2000, MNRAS, 313, 364
\bibitem[]{} Petitjean P., Surdej J., Smette A., Shaver P., M\"ucket
J., Remy M., 1998, A\&A, 334, L45 
\bibitem[]{} Porciani C., Matarrese S., Lucchin F., Catelan P., 1998,
MNRAS, 298, 1097
\bibitem[]{} Rauch M., 1998, ARA\&A, 36, 267
\bibitem[]{} Rauch M., Haehnelt M., 1995, MNRAS, 275L, 76R
\bibitem[]{} Rauch M., Sargent W.L.W. Barlow, T.A., 1999, ApJ, 515, 500
\bibitem[]{} Rauch M., Miralda-Escude J., Sargent W. L. W., Barlow
T. A., Weinberg D. H., Hernquist L., Katz N., Cen R., Ostriker
J. P., 1997, ApJ, 489, 7 
\bibitem[]{} Reisenegger A., Miralda-Escud\'e J., 1995, ApJ, 449, 476
\bibitem[]{} Roy Choudhury T., Padmanabhan T., Srianand R., MNRAS, 2001, 322, 561
\bibitem[]{} Roy Choudhury T., Srianand R., Padmanabhan T., 2001, 
MNRAS, submitted, preprint astro-ph/00012498 
\bibitem[]{} Schaye J., Theuns T., Rauch M., Efstathiou G., 
Sargent W.L.W., MNRAS, 2000, 318, 817
\bibitem[]{} Smette A., Robertson J.G., Shaver P.A., Reimers D.,
Wisotzki L., K\"ohler Th., 1995, A\&AS, 113, 199
\bibitem[]{} Smette A., Surdej J., Shaver P.A., Foltz C.B., 
Chaffee F.H., Weymann R.J., Williams R.E., Magain P., 1992, ApJ, 389, 39 
\bibitem[]{} Sugiyama N., 1995, ApJS, 100, 281 
\bibitem[]{} Theuns T., Schaye J., Haehnelt M.G., 1999, MNRAS,
submitted, preprint astro-ph/9908288  
\bibitem[]{} Theuns T., Leonard A., Efstathiou G., Pearce F.R., 
Thomas P.A., 1998, MNRAS, 301, 478
\bibitem[]{} Williger G.M., Smette A., Hazard C., Baldwin J.A.,
McMahon R.G., 2000, ApJ, 532, 77
\bibitem[]{} White M., Croft R.A.C., 2000, preprint astro-ph/0001247
\bibitem[]{} Young P. A., Impey C. D., Foltz, C. B., 2000, preprint 
astro-ph/0010058
\bibitem[]{} Zel'dovich ya. B., 1970, A\&A, 5, 84
\bibitem[]{} Zhang Y., Anninos P., Norman M.L., 1995, ApJ, 453, L57
\bibitem[]{} Zhang Y., Anninos P., Norman M.L., Meiksin A., 1997, ApJ, 485, 496 
\bibitem[]{} Zhang Y., Meiksin A., Anninos P., Norman M.L., 1998, ApJ, 495, 63 
~\end{thebibliography}
\end{document}